\documentclass[12pt]{article}
\usepackage[font=scriptsize,labelfont=bf]{caption}

\usepackage{bbm}

\usepackage{natbib}

\usepackage{charter}

\usepackage{float}
\usepackage[font=scriptsize]{caption}
\usepackage{appendix}
\usepackage{titlesec}
\usepackage{mathrsfs}
\usepackage{amsmath}
\usepackage{amssymb}
\usepackage{amsthm}
\usepackage{mathrsfs}
\usepackage{amssymb}
\usepackage{setspace}
\usepackage{tikz}
\usepackage{xcolor}
\usepackage{enumitem}
\setlist{nolistsep}

\colorlet{drkblue}{blue!61.8!black}

\usepackage[bookmarks=false]{hyperref}
\usepackage{microtype}
\usepackage{bookmark}
\hypersetup{
     pdfborder = {0 0 0},
  colorlinks=true,
 citecolor=drkblue,
linkcolor=drkblue,
urlcolor=drkblue,
pdfstartview={XYZ null null 0.80}
}

\usepackage[hyperpageref]{backref}
\hypersetup{hypertexnames=false}
\microtypesetup{protrusion=false}

\usepackage[left=2.5cm,top=3.5cm,bottom=3.5cm,right=3cm]{geometry}
\usepackage{setspace}
\titlespacing*{\section}{0pt}{.8\baselineskip}{0.5\baselineskip}
\titlespacing*{\subsection}{0pt}{.6\baselineskip}{0.5\baselineskip}

\newtheorem*{definition}{Definition}
\newtheorem{theorem}{Theorem}
\newtheorem{corollary}{Corollary}
\newtheorem{lemma}{Lemma}
\newtheorem{proposition}{Proposition}

\newtheorem{example}{Example}

\newcommand{\C}{\mathcal{C} }

\newcommand{\B}[3]{#1\mathrel{\mathcal{B_\mu}}\{#2,#3\}}

\usepackage{tikz,array} 
\usetikzlibrary{positioning,automata}
\usetikzlibrary{arrows,automata}
\usetikzlibrary{arrows, quotes}
\usepackage{color, colortbl}
\usetikzlibrary{decorations.pathreplacing}

\usetikzlibrary{fit}
\usetikzlibrary{calc}

\newsavebox{\boxA}
\newsavebox{\boxB}
\newlength{\lenA}
\newlength{\lenB}

\definecolor{Gray}{gray}{0.9}
\newcommand{\X}{\Omega}

\newcommand{\rcf}{\texttt{PCF}}

\definecolor{greenA}{RGB}{160, 210, 160}  % aç?k ama art?k pastel de?il
\definecolor{greenB}{RGB}{95, 170, 120}   % dengeli aç?k-orta
\definecolor{greenC}{RGB}{45, 125, 85}    % net orta koyuluk
\definecolor{greenD}{RGB}{20, 75, 50} 

\definecolor{burgundyA}{RGB}{122, 24, 45}

\definecolor{turquoiseA}{RGB}{35, 150, 155}

\title{\textbf{Self-ordered choice models}\thanks{
I am grateful to Christopher Chambers, Battal Dogan, Faruk Gul and Ariel Rubinstein for helpful discussions and feedback. I thank seminar participants at University of Maryland, Georgetown University, Duke University, Princeton University, and participants at several conferences and workshops for valuable comments and suggestions.\@ This project has received funding from the European Union's Horizon 2020 research and innovation programme under the Marie Sk{\l}odowska-Curie grant agreement No.~837702.\@ 
\newline First posted version: December 27, 2022.   
  }}%\thanks{We are grateful to Ariel Rubinstein, Vicki Knoblauch, Efe Ok, Yusufcan Masatlioglu, Erkut Ozbay, Ed Green, Kfir Eliaz, Mustafa Celebi Pinar, Miguel Ballester, Ran Spiegler, Burak Pac, Paola Manzini, Marco Mariotti, Herve Moulin, Michael Richter,  Battal Dogan, Selman Erol, Faruk Gul, Wolfgang Pesendorfer, Leeat Yariv, David Dillenberger, Pietro Ortoleva, and seminar participants at Bilkent University, Princeton University, University of Maryland, University of Bristol, University of Glasgow, ADA University, New School of Economics, and Higher School of Economics, the editor and the referees for their helpful comments and suggestions.\@ 
\usepackage{hyperref}

\author{\large{K}\small{EMAL} \large{Y}\small{ILDIZ}\thanks{Department of Economics, Bilkent University. Email: \texttt{kemal.yildiz@bilkent.edu.tr}}
}

\begin{document}

    \maketitle\ 
  \vspace*{-2cm}      
   \begin{abstract}
  \vspace*{-.5cm}      
   
 \noindent

The same aggregate choice data can reflect different patterns of
individual behavior, with different implications for economic analysis.
We study when ranking individuals by their choices can resolve this
ambiguity while preserving a model's behavioral assumptions. We call
a model \emph{self-ordered} if every aggregate choice pattern it explains
can also be explained by individuals who behave as the model permits
and can be consistently ranked by their choices. We show that this
holds precisely when, for any two allowed patterns of behavior, both
the pattern formed by taking the higher-ranked choice in every
situation and the pattern formed by taking the lower-ranked choice in
every situation are also allowed by the model. This provides
a practical way to assess models and to restrict or extend them when
necessary. For rational choice, we identify the smallest enlargement
of the model with this property. Finally, we establish conditions
under which the underlying ranking can itself be recovered from the
behavior allowed by the model in an essentially unique way.

%The same aggregate choice data can reflect different patterns of
%individual behavior, with different implications for economic analysis.
%We study when ordering individuals by their choices can resolve this
%ambiguity while preserving the behavioral assumptions of a given model.
%We call a choice model
%\emph{self-ordered} if every probabilistic choice function it generates
%admits an ordered representation using only types from the model. We
%characterize self-ordered choice models by the requirement that their
%admissible types form a lattice under the relevant ordering. This
%characterization provides an operational criterion for determining
%whether ordered identification is compatible with a prespecified model
%and a recipe for restricting or extending models to achieve this
%compatibility. Applying the result to rational choice, we identify
%exactly which additional behavioral types are needed to obtain a unique
%ordered representation.
%Finally, we establish conditions under which the underlying ordering can be
%recovered essentially uniquely from the behavior admitted by the model.


             \end{abstract}  
   
%explained via a structurally-invariant model.
%               any deterministic or stochastic choice behavior can be modeled.
              
%               \noindent \textit{JEL} Classification Numbers: D01, D11, D71.      

        \newpage
        \doublespacing

\vspace*{-2cm}      
               \tableofcontents
                
                \newpage

 \section{Introduction}

 An economist studying demand often observes market-level choice frequencies rather than individual consumers' choices. For example, she may observe the shares of consumers choosing different mobile-phone plans as the set of available plans changes. These aggregate choice frequencies reflect heterogeneity in the population: different consumers may follow different decision rules even when they face the same alternatives. We refer to data specifying how frequently each alternative is chosen from each menu as probabilistic choice.

To use such aggregate data for counterfactual, comparative-static, or welfare
analysis, the economist would like to recover the underlying heterogeneity. A
standard approach is to posit a set of deterministic behavioral types and
interpret the observed choice frequencies as arising from a distribution over
these types. In a demand setting, these types describe consumers' decision
rules. Under the \textit{random utility model}, for example, each type chooses
its most preferred alternative from each menu, and aggregate choice frequencies
reflect differences in preferences across consumers. Recovering the
distribution of types does more than match observed market shares: it describes
how behavior varies across consumers and provides a basis for subsequent
economic analysis.

The difficulty is that observed probabilistic choice typically does not uniquely determine the underlying distribution of types. Different distributions can generate the same choice frequencies while describing different patterns of heterogeneity in the population. This underidentification matters economically: representations that fit the same data may lead to different conclusions in subsequent economic analysis. Thus, explaining observed choice frequencies does not necessarily identify the heterogeneity that generates them. A natural question is whether additional structure can resolve this ambiguity while preserving the behavioral restrictions of the underlying economic model.

 One way to resolve this ambiguity is to require the behavioral types used to represent the population to be ordered. Given an ordering of alternatives within each menu, choice types can be compared according to whether they consistently select higher- or lower-ranked alternatives. \cite{prc} show that whenever a probabilistic choice function admits a representation whose supporting types are ordered in this way, that representation is unique. Requiring the supporting types to be ordered can therefore resolve the underidentification of population heterogeneity without imposing parametric restrictions. However, an important issue remains for economic applications: the types needed for this unique ordered representation may fall outside the analyst's chosen model and thus violate its behavioral restrictions.

 We call a choice model \emph{self-ordered} when this problem does not arise: every probabilistic choice function that can be represented using types from the model also admits a unique ordered representation using types from the same model. Being self-ordered therefore allows the analyst to resolve underidentification by requiring the supporting types to be ordered, without relaxing the model's behavioral restrictions. Our main result (Theorem~\ref{progressiveifflattice}) provides a simple characterization of this property: a choice model is self-ordered if and only if its choice types form a \emph{lattice} under the relevant ordering.

The lattice requirement has a direct behavioral interpretation. For any two admissible types, the model must also contain the type that selects the higher-ranked of their choices from each menu and the type that selects the lower-ranked choice.  The resulting types are the \emph{join} and \emph{meet} of the original pair, respectively. Thus, being self-ordered requires the model to be closed under these menu-by-menu combinations of its own behavioral types.

The lattice characterization provides a precise recipe for restricting or extending familiar choice models to make them self-ordered. An important case is the rational choice model, which is not self-ordered. Consequently, some probabilistic choice functions generated by the random utility model have unique ordered representations that necessarily include nonrational choice types. Requiring the supporting types to be ordered can therefore take the analyst outside the rational choice model. This raises a natural question: which additional choice types are needed to make the model self-ordered, and do they have an economically meaningful interpretation?

We answer this question by characterizing the \emph{minimal self-ordered
extension} of the rational choice model (Theorem~\ref{minimalextension}). Its
two characterizing axioms have clear behavioral interpretations. 
\textit{Inferior-removal monotonicity} requires that removing an alternative
ranked below the original choice cannot lead to a lower-ranked choice. \textit{Superior-addition monotonicity} requires that adding an
alternative ranked above the choice from the expanded menu cannot lead to a
choice ranked below the choice from the original menu.

The underlying primitive ordering is at the heart of the analysis, as it
captures the economic dimension along which alternatives are compared. For
example, lotteries may be ranked by their expected monetary payoffs,
consumption bundles by their simplicity, policies by their associated tax
revenue, or products by their carbon emissions. In some applications, this
ordering is naturally given by the economic environment; in others, it must be
identified. The minimal self-ordered extension characterized in
Theorem~\ref{minimalextension} provides a setting in which to address this
question. Under a natural \textit{richness} condition,
Theorem~\ref{revealed} gives necessary and sufficient conditions for recovering
the underlying primitive ordering from the choice model and shows that this
ordering is essentially unique. The result shows that the ordered
interpretation of heterogeneity need not rest on an externally imposed ranking:
it can instead be grounded in the behavioral restrictions embodied in the
model itself.

We also examine broader implications of the framework. The similarity-based
choice model in Example~\ref{similarity} illustrates how the lattice structure
accommodates multiple behavioral characteristics. The examples in
Section~\ref{Section3.3} connect the framework to different economic
environments and further extend its scope. Finally,
Proposition~\ref{universal} uses a simple closure condition, stronger than the
lattice requirement, to characterize the choice models that are self-ordered
under every specification of the primitive ordering. Together, these results
illustrate the breadth of the framework and the role of the primitive ordering
in determining whether a model is self-ordered.

%we examine whether the existence and uniqueness of ordered representations alone can justify requiring comparisons of choice types to be generated by primitive orderings. 

%: if every probabilistic choice function admits a unique ordered representation, then the comparison relation must be induced by a family of primitive orderings, one for each menu.
%We conclude with 

The following example illustrates what it means for a model to be self-ordered and how the lattice characterization in Theorem~\ref{progressiveifflattice} applies.

\begin{example}
\label{ex1}
\normalfont
Consider a population of agents choosing among electric ($e$), hybrid ($h$), and gasoline-powered ($g$) cars. The same agents also choose a writing instrument from among a fountain pen filled from bottled ink ($fi$), a cartridge fountain pen ($fc$), and a disposable pen ($d$). The analyst observes the population's aggregate choice behavior, summarized by the probabilistic choice function $\rho$ shown in the left panel of Figure~\ref{f1}.

The analyst's primary interest is the eco-friendliness of choices. Accordingly, she ranks cars as $e>h>g$ and writing instruments as $fi>fc>d$, with higher-ranked alternatives viewed as more environmentally friendly.\footnote{Because the framework allows arbitrary menu domains, alternatives that never appear together in a menu need not be compared. In this example, no comparison between cars and writing instruments is required.}
The analyst considers a choice model $\mu$ to explain the observed $\rcf$. The right panel of Figure~\ref{f1} depicts the four choice types that constitute $\mu$. These types are partially ordered by the comparison relation $\vartriangleright$, induced by $>$, which compares the eco-friendliness of their choices across the two menus. In particular, $c_1 \vartriangleright c_2 \vartriangleright c_4$ and $c_1 \vartriangleright c_3 \vartriangleright c_4$.

\begin{figure}[ht]
        \savebox{\boxA}
        {
                \scalebox{1.2}{

                        %       \begin{center}
                        \begin{tabular}{c|cccc}
                                $\rcf$ $\rho$ & $e$ & $h$ & $fi$ & $d$\\
                                \cline{1-5}

         $\{e,h,g\}$ & $\frac{2}{3}$ & $\frac{1}{3}$ & $-$ & $-$\\ 
                                $\{fi,fc,d\}$ & $-$ & $-$ & $\frac{2}{3}$ & $\frac{1}{3}$\\
                        \end{tabular}
                        
                        %       \end{center}

                }
                
        }
        \settoheight{\lenA}{\usebox{\boxA}}
        
        \savebox{\boxB}
        {        \scalebox{0.6}{ \begin{tikzpicture}
                        
                        %2th layer

                        \node (l21) at (0,0) {\textcolor{greenD}{$\boxed{\mathbf{c_1=[e,fi]}}$}};
                        \node(21) at (0,-0.5) {};
                        \draw [fill = black](21) circle (4pt);

                        %3th layer

                        \node (l31) at (-3.5,-2) {\textcolor{greenC}{$\boxed{\mathbf{c_2=[h,fi]}}$}};
                        \node(31) at (-3,-2.5) {};
                        \draw [fill = black](31) circle (4pt);

                        \node (l33) at (3.5,-2) {\textcolor{greenB}{$\boxed{\mathbf{c_3=[e,d]}}$}};
                        \node(33) at (3,-2.5) {};
                        \draw [fill = black](33) circle (4pt);
                        
                        %4th layer

                        \node (l42) at (0,-5) {\textcolor{greenA}{$\boxed{\mathbf{c_4=[h,d]}}$}};
                        \node(42) at (0,-4.5) {};
                        \draw [fill = black](42) circle (4pt);

                        \draw [dashed, color = black, thick] (31) -- (42) ;

                        \draw [dashed, color = black, thick]  (21) -- (31);

                        \draw [dashed, color = black, thick] (21) -- (33);

                        \draw [dashed, color = black, thick] (33) -- (42);

                        \node (ll81) at (-5,-0) {\textcolor{turquoiseA}{\textbf{ $\frac{1}{3}c_1 \oplus \frac{1}{3}c_2 \oplus \frac{1}{3} c_3$}}};
                        \draw [->,>=stealth',shorten >=1pt,auto,node distance=2.8cm,
                        thick, scale=0.8, color=turquoiseA, transform shape] (ll81) -- (21);

                        \draw [->,>=stealth',shorten >=1pt,auto,node distance=2.8cm,
                        thick, scale=0.8, color=turquoiseA, transform shape] (ll81) -- (31);

                        \draw [->,>=stealth',shorten >=1pt,auto,node distance=2.8cm,
                        thick, scale=0.8, color=turquoiseA, transform shape] (ll81) -- (33);

                        %                                       %\pause  

                        \node (ll82) at (5,-5) {\textcolor{burgundyA}{\textbf{ $\frac{2}{3}c_1 \oplus  \frac{1}{3} c_4$}}};
                        \draw [->,>=stealth',shorten >=1pt,auto,node distance=2.8cm,
                        thick, scale=0.8, color=burgundyA, transform shape] (ll82) -- (42); 
                        
                        \draw [->,>=stealth',shorten >=1pt,auto,node distance=2.8cm,
                        thick, scale=0.8, color=burgundyA, transform shape] (ll82) -- (21);

                        \end{tikzpicture}
        }}
        
        \settoheight{\lenB}{\usebox{\boxB}}
        \noindent
        \hfil\usebox{\boxA}
        \hfil\raisebox{-0.45\lenB}{\usebox{\boxB}}\hfil\par
\caption{Each node in the graph on the right represents a choice type and specifies its choices from the menus $\{e,h,g\}$ and $\{fi,fc,d\}$. Dotted lines indicate comparisons under $\vartriangleright$, the relation induced by $>$. For example, $c_1\vartriangleright c_2$ because both types choose $fi$ from $\{fi,fc,d\}$, while $c_1$ chooses $e$ from $\{e,h,g\}$ and $c_2$ chooses the lower-ranked alternative $h$.\label{f1}}      \end{figure}

The observed $\rcf$ can be represented by a probability distribution that assigns equal weight to the choice types $c_1$, $c_2$, and $c_3$. This representation is not ordered because $c_2$ and $c_3$ are incomparable: $c_2$ chooses the less eco-friendly car $h$ but the more eco-friendly pen $fi$, whereas $c_3$ chooses the more eco-friendly car $e$ but the less eco-friendly pen $d$.
Relative to $c_3$, we can interpret $c_2$ as more sensitive to plastic waste but less sensitive to carbon emissions. These two types thus capture heterogeneity in environmental sensitivity along two distinct dimensions, rather than a single ranking of overall eco-friendliness.

Alternatively, the same $\rcf$ can be represented by assigning weight $\frac{2}{3}$ to $c_1$ and weight $\frac{1}{3}$ to $c_4$. Type $c_1$ selects the most eco-friendly alternative from each menu, whereas $c_4$ selects the lower-ranked alternatives $h$ and $d$. Since $c_1\vartriangleright c_4$, this representation is ordered. Although the observed $\rcf$ admits infinitely many other representations, none of them is ordered. The existence and uniqueness of this ordered representation follow from Theorem~1 of \cite{prc}.

We now ask whether the model $\mu$ is self-ordered: does every $\rcf$ generated by a mixture of types in $\mu$ also admit an ordered representation using only types in $\mu$?
%\footnote{Equivalently, each representation of the former kind (highlighted in blue) would need to be paired with a corresponding ordered representation (highlighted in red).} 
The observed $\rho$ satisfies this requirement: it is generated by a mixture of $c_1$, $c_2$, and $c_3$, and its ordered representation uses $c_1$ and $c_4$, both of which belong to $\mu$. This alone does not establish that $\mu$ is self-ordered, since the requirement must hold for every $\rcf$ that the model can generate.
By contrast, removing $c_4$ from $\mu$ would leave a model that still represents $\rho$ but cannot provide an ordered representation of it. The reduced model would therefore not be self-ordered.

Restricting $\mu$ to the choice types $c_1$ and $c_4$ would preserve the ordered representation of $\rho$. This two-type model forms a \textit{chain lattice}, in which all choice types are comparable. Admitting $c_2$ and $c_3$ expands the set of probabilistic choice patterns that the model can explain. These two types are incomparable because their relative environmental sensitivity differs across carbon emissions and plastic waste. The four-type model therefore accommodates multidimensional heterogeneity, whereas the two-type model permits only a common ranking of types across both menus.

 Establishing that $\mu$ is self-ordered by direct inspection would require checking that every $\rcf$ generated by a mixture of $c_1$, $c_2$, $c_3$, and $c_4$ admits an ordered representation within $\mu$. Theorem~\ref{progressiveifflattice} avoids the need for these checks: since $\langle \mu, \vartriangleright\rangle$ forms a lattice, $\mu$ is self-ordered.
\end{example}

\vspace{-1cm}
\subsection{Related literature}
\label{lit}
\vspace{-.5cm}
We develop a general approach to addressing the underidentification
that commonly arises in probabilistic choice models.\@
Recent studies that impose an ordering on the population motivate
our definition of a self-ordered choice model.\@
This subsection situates our approach within that literature and
clarifies the conceptual significance and economic relevance of
our contribution.

Ordered probabilistic choice has roots in the discrete choice literature
(e.g., \citealp{amemiya1981qualitative, small1987discrete,
Agresti1984}). This literature studies how individuals make
probabilistic choices among exogenously ordered alternatives.
More recently, \citet*{scrum} have reinvigorated interest in
ordered probabilistic choice, giving rise to a growing body of work
\citep*{barseghyan2021discrete, tserenjigmid2021order,
turansick2022, prc, apesteguia2023random,
apesteguia2023testable, petri2023binary,
masatlioglu2024growing}. 

Our contribution departs from this literature by focusing on the
structure of choice models rather than on any particular model.
We ask which structural properties guarantee a unique ordered
representation for every probabilistic choice behavior that a
model can explain.

\citet*{scrum} are the first to impose an ordering on the population
to refine the random utility model and obtain a unique representation.\@
Alongside their axiomatic characterization, they show that when a RUM
is represented as a probability distribution over rational choice types
that are comparable under this ordering---a \textit{single crossing
random utility model} (SCRUM)---the representation is unique.\footnote{See \citet*{costa2020single} for an extension to
probabilistic choice correspondences.}
Building on this observation, \citet{prc} show that every probabilistic
choice function admits a unique representation as a probability
distribution over mutually comparable choice types.\@
\cite*{cmy} connect ordered probabilistic choice to copula theory and
show that this unique ordered representation is precisely the one
generated by the Fréchet--Hoeffding upper bound, or comonotonic copula.
Together, these results motivate using an ordering to select a single
representation from the typically infinite set of possible
representations, as illustrated in Example~\ref{ex1}.
We instead ask when the unique ordered representation is supported
entirely on choice types admitted by a prespecified behavioral model.

In our terminology, \citet{prc} establish that one particular
model---the model containing all choice types---is self-ordered.\@
Our aim is to characterize the entire class of self-ordered choice
models, a question of both conceptual and technical significance.

We next discuss why this generality matters economically.
Both \citet*{scrum} and \citet{prc} present examples in which
agents' choices are ordered according to a single characteristic.\@
It remains unclear, however, whether ordered representations exist
for models in which agents' choices vary along multiple behavioral
characteristics. The equivalence between self-ordered models and lattices
(Theorem \ref{progressiveifflattice}) suggests that such
characteristics can be specified separately while preserving
a unique ordered representation. Specifically, the similarity-based
choice model in Example~\ref{similarity} illustrates an economically
relevant form of multidimensional heterogeneity.

Section~\ref{Section3.3} shows that the framework applies to several
economically relevant environments, including monotone comparative
statics, competitive equilibrium prices, and stable matchings.
In each environment, standard economic primitives induce a natural
ordering over the relevant objects, and existing results or a direct
argument establish that the admissible objects form a lattice.
Our characterization therefore implies that aggregate observations
consistent with these models admit unique ordered representations.
Under the ordering restriction, these representations allow the analyst
to identify consumers' switching patterns across observed menus, the
co-movement of equilibrium prices across goods, and the joint distribution
of matching outcomes, while preserving the restrictions of the underlying
economic models.

        \vspace{-1cm}
\section{Self-ordered choice models}
        \label{model}
\vspace{-.5cm}
Let $X$ be the \textbf{alternative set}, containing $n$ elements.\@
A \textbf{menu} $S$ is a subset of $X$ containing at least two
alternatives.\@ The \textbf{choice domain} $\X$ is a nonempty
collection of menus, allowing the available data to cover only
some menus (limited data sets).\@
A \textbf{choice type} specifies one alternative from each menu in
the choice domain; formally, it is a mapping $c:\X \rightarrow X$
such that $c(S)\in S$ for every $S\in\X$.\@
We denote the set of all choice types by $\C$.\@
A \textbf{choice model} $\mu$ is
a nonempty set of choice types.\@ Choice procedures with different
algorithmic formulations generate the same choice model if they are
observationally indistinguishable in revealed-preference terms---that
is, if they rationalize the same set of choice types.\@

A \textbf{probabilistic choice function} ($\rcf$) $\rho$ assigns
each menu $S\in\X$ a probability distribution over its alternatives.\@
We denote by $\rho_x(S)$ the probability that alternative $x$ is
chosen from menu $S$.\@ Every $\rcf$ can be represented by a
probability distribution over choice types whose induced choice
probabilities coincide with $\rho$, but this representation need
not be unique.\@ For a choice model $\mu$, the associated
\textbf{probabilistic choice model} $\Delta(\mu)$ is the set of all
$\rcf$s that admit a representation using only choice types in
$\mu$.
For example, a choice type is \textbf{rational} if its choices can
be rationalized by a strict preference relation---that is, if it
selects the most-preferred alternative from every menu.\@ A $\rcf$
is a \textbf{random utility model} (RUM) if it admits a representation
as a distribution over rational choice types.

For each menu $S\in\X$, a \textbf{primitive ordering} $>_S$ ranks the alternatives in $S$ and is complete, transitive, and asymmetric. We write $\geq_S$ for the associated weak ordering: $x\geq_S y$ if $x>_S y$ or $x=y$. These menu-specific rankings induce a \textbf{comparison relation} $\vartriangleright$ over choice types. Specifically, $c\vartriangleright c'$ if $c$ selects an alternative that is (weakly) higher-ranked than the alternative selected by $c'$ from every menu and strictly higher-ranked from at least one menu; that is,
$c(S)\geq_S c'(S)$ for every $S\in\X$, with $c(S)\neq c'(S)$ for some $S\in\X$. The resulting relation $\vartriangleright$ is a partial order over choice types. We write $c\trianglerighteq c'$ if $c\vartriangleright c'$ or $c=c'$.

Notably, we allow primitive orderings to vary across menus.\@
Together with our allowance for limited data sets, this permits
incomparabilities across the full set of alternatives. Example
\ref{ex1} illustrates such a case: the alternatives in two menus
are of entirely different kinds---cars and pens---and therefore
need not be compared.\@ Menu-dependent primitive orderings also
allow us to capture context-dependent forces, such as temptation
or information-processing costs, whose effects depend on the
presence of more tempting or memorable alternatives in the menu.
We now introduce the central notion of our analysis.

\begin{definition}[Self-Ordered Choice Model]
%\normalfont
Let $\vartriangleright$ be the partial order over choice types
induced by the primitive orderings $\{>_S\}_{S\in\X}$.\@ A choice
model $\mu$ is \textbf{self-ordered} with respect to
$\vartriangleright$ if every $\rcf$ $\rho\in\Delta(\mu)$ admits a
representation in which all choice types receiving positive
probability are ordered by $\vartriangleright$. Formally, $\rho$
can be represented as a probability distribution over some
$\{c^1,\ldots,c^k\}\subset\mu$ such that
$c^1\vartriangleright c^2\vartriangleright\cdots
\vartriangleright c^k$.
\end{definition}

\citet{prc} introduce the notion of an \textbf{ordered
representation}: a $\rcf$ has an ordered representation if it admits
a representation as a probability distribution over choice types
$\{c^1,\ldots,c^k\}$ such that
$c^1\vartriangleright c^2\vartriangleright\cdots
\vartriangleright c^k$.\@ These choice types need not belong to any
prespecified choice model $\mu$.\@ By contrast, a self-ordered choice
model ensures that every probabilistic choice it can explain has an
ordered representation using only model-admissible types; as discussed
in the introduction, this representation is unique.

To see why an ordered representation is unique whenever it exists,
consider its $\vartriangleright$-highest choice type $c^1$.\@ For
each menu $S\in\X$, this type selects the $>_S$-highest alternative
that $\rho$ chooses with positive probability. Thus, $c^1$ is
determined by $\rho$, and its probability weight is the smallest
choice probability assigned to $c^1(S)$ across menus $S$.\@ After
removing the contribution of $c^1$, the same argument identifies
$c^2$ and its probability weight. Repeating this procedure uniquely
determines the entire ordered representation.\@ Consequently, the
same $\rcf$ $\rho$ may belong to both $\Delta(\mu)$ and
$\Delta(\mu')$ for distinct models $\mu$ and $\mu'$, but its ordered
representation, whenever it exists, is the same in both models.

%In contrast, changing the primitive orderings would alter the induced comparison relation and, consequently, the ordered representation itself.

\subsection{Equivalence between self-ordered choice models  and lattices }
% ordered choice models and lattices
% ordered choice models are the ones that have lattice structure

\label{equivalence}

In this section, we characterize the entire class of self-ordered
choice models. Fix the primitive orderings $\{>_S\}_{S\in\X}$ and
let $\vartriangleright$ be the associated partial order over choice
types.\@ To understand the relevant structure, consider combining
any two choice types $c$ and $c'$ menu by menu. One combination
retains the $>_S$-higher-ranked choice from each menu $S$, while the
other retains the $>_S$-lower-ranked choice. These combinations form
the natural upper and lower envelopes of the two types and allow us
to ask whether a model contains the extremal types generated by its
own members.\@ Formally, the first combination is the \textit{join}
$c\vee c'$, defined by selecting the $>_S$-higher-ranked alternative
among $c(S)$ and $c'(S)$ from every menu $S$. The second is the
\textit{meet} $c\wedge c'$, defined analogously using the
$>_S$-lower-ranked alternative. In order-theoretic terms, the join
is their least upper bound and the meet is their greatest lower bound
under $\trianglerighteq$.

\begin{definition}[Lattice]
The pair $\langle\mu,\vartriangleright\rangle$ is a
\textbf{lattice} if the choice model $\mu$ is closed under joins
and meets: for every $c,c'\in\mu$, both $c\vee c'$ and
$c\wedge c'$ also belong to $\mu$.
\end{definition}

Thus, throughout the paper, a lattice of choice types means a
sublattice of the ambient lattice $\langle\C,\vartriangleright\rangle$,
with joins and meets computed pointwise.
The following theorem provides our main structural characterization.
It shows that closure under these two menu-by-menu operations is
exactly the condition required for a choice model to be self-ordered.

\begin{theorem}[Lattice Equivalence]
\label{progressiveifflattice}
Let $\mu$ be a choice model, and let $\vartriangleright$ be the
partial order over choice types induced by the primitive orderings
$\{>_S\}_{S\in\X}$.\@ Then, $\mu$ is self-ordered with respect to
$\vartriangleright$ if and only if
$\langle\mu,\vartriangleright\rangle$ is a lattice.
\end{theorem}
The \textit{only if} direction can be understood through a simple
mixture. Take any $c,c'\in\mu$ and let $\rho$ be the $\rcf$ generated
by assigning equal probability to these two choice types. Because
$c,c'\in\mu$, we have $\rho\in\Delta(\mu)$.\@ The unique ordered
representation of $\rho$ assigns equal probability to $c\vee c'$
and $c\wedge c'$.\@ If $\mu$ is self-ordered, this representation
must use only choice types in $\mu$. Hence, $c\vee c'\in\mu$ and
$c\wedge c'\in\mu$, so $\mu$ is closed under joins and meets.

For the converse, suppose that
$\langle\mu,\vartriangleright\rangle$ is a lattice and fix a
$\rcf$ $\rho\in\Delta(\mu)$.\@ We first construct the unique ordered
representation of $\rho$ and then show that every choice type in its
support belongs to $\mu$.

\noindent\textbf{Step 1 (Order and stack the probabilities).}
For each menu $S$, let $\rho^+(S)=\{x\in S:\rho_x(S)\neq 0\}$ and list its elements as
$x_1^S>_S\cdots>_S x_{m_S}^S$. Partition $(0,1]$ into consecutive half-open intervals,
one for each $x\in\rho^+(S)$, arranged from left to right according to this ordering.
Write $I_{Sx}=(l_{Sx},u_{Sx}]$ and assign it length
$u_{Sx}-l_{Sx}=\rho_x(S)$.

\begin{figure}[!ht]
\centering
\begin{tikzpicture}[x=.82cm,y=1cm,font=\small]

    % Interval bars
    \draw[fill=gray!8]  (0,2.20) rectangle (1.70,2.70);
    \draw[fill=blue!14] (1.70,2.20) rectangle (4.70,2.70);
    \draw[fill=gray!8]  (4.70,2.20) rectangle (6.20,2.70);
    \draw[fill=gray!8]  (6.20,2.20) rectangle (8.20,2.70);
    \draw[fill=gray!8]  (8.20,2.20) rectangle (10,2.70);

    \draw[fill=gray!8]  (0,.55) rectangle (1.20,1.05);
    \draw[fill=blue!14] (1.20,.55) rectangle (4.40,1.05);
    \draw[fill=gray!8]  (4.40,.55) rectangle (6.80,1.05);
    \draw[fill=gray!8]  (6.80,.55) rectangle (8.15,1.05);
    \draw[fill=gray!8]  (8.15,.55) rectangle (10,1.05);

    % Lengths used later in the proof
    \draw[blue!65!black,thick,|-|] (1.70,3.02) -- (10,3.02);
    \draw[blue!65!black,thick,|-|] (0,-.04) -- (4.40,-.04);

    % Common uniform draw
    \draw[red!70!black,dashed,very thick]
        (3.55,-.20) -- (3.55,3.38);

    % Labels for menu S
    \node[anchor=east] at (-.35,2.45) {$S:$};
    \node at (.85,2.45) {$I_{Sw}$};
    \node[fill=blue!14,inner sep=1pt] at (3.20,2.45) {$I_{Sx}$};
    \node at (5.45,2.45) {$I_{Sy}$};
    \node at (7.20,2.45) {$\cdots$};
    \node at (9.10,2.45) {$I_{Sz}$};

    \node[anchor=north] at (0,2.14) {$0$};
    \node[anchor=north] at (10,2.14) {$1$};

    \node[fill=white,inner sep=1pt] at (5,1.68)
        {$w>_S x>_S y>_S\cdots>_S z$};

    \node[anchor=west,blue!60!black] at (10.35,2.45)
        {$c_r(S)=x$};

    \node[
        fill=white,
        inner sep=1pt,
        text=blue!65!black,
        above
    ] at (5.85,3.02) {$1-l_{Sx}$};

    % Labels for menu S'
    \node[anchor=east] at (-.35,.80) {$S':$};
    \node at (.60,.80) {$I_{S'w'}$};
    \node[fill=blue!14,inner sep=1pt] at (2.80,.80) {$I_{S'x'}$};
    \node at (5.60,.80) {$I_{S'y'}$};
    \node at (7.48,.80) {$\cdots$};
    \node at (9.08,.80) {$I_{S'z'}$};

    \node[anchor=north] at (0,.49) {$0$};
    \node[anchor=north] at (10,.49) {$1$};

    \node[fill=white,inner sep=1pt] at (5,-.48)
        {$w'>_{S'}x'>_{S'}y'>_{S'}\cdots>_{S'}z'$};

    \node[anchor=west,blue!60!black] at (10.35,.80)
        {$c_r(S')=x'$};

    \node[
        fill=white,
        inner sep=1pt,
        text=blue!65!black,
        above
    ] at (2.20,-.04) {$u_{S'x'}$};

    % Label for the common draw
    \node[
        text=red!70!black,
        fill=white,
        inner sep=1pt,
        above
    ] at (3.55,3.38) {$r$};

\end{tikzpicture}

\caption{Uniform decomposition. Each row partitions $(0,1]$ into intervals whose
lengths are the menu-specific choice probabilities and orders them from higher- to
lower-ranked alternatives. The same uniform draw $r$ selects one interval in every
menu; here, $c_r(S)=x$ and $c_r(S')=x'$. The blue lines indicate the lengths
$1-l_{Sx}$ and $u_{S'x'}$ used in the proof.}
\label{intervals}
\end{figure}

\noindent\textbf{Step 2 (Use a common quantile).}
Draw a single number $r$ uniformly from $(0,1]$ and use the same draw for every menu.
Define the choice type $c_r$ by setting $c_r(S)=x$ if and only if $r\in I_{Sx}$.
For every menu $S$ and alternative $x$, we then have
$\Pr(c_r(S)=x)=|I_{Sx}|=\rho_x(S)$, so the distribution induced by $r$ represents
$\rho$. Moreover, if $r<r'$, then $c_r(S)\geq_S c_{r'}(S)$ for every menu $S$ and,
hence, $c_r\trianglerighteq c_{r'}$. The inequality is strict whenever
$c_r\neq c_{r'}$. Thus, after listing the distinct choice types in the support from
highest to lowest, the support can be written as $\{c^i\}_{i=1}^k$, where
$c^1\vartriangleright c^2\vartriangleright\cdots\vartriangleright c^k$.\footnote{See
\citet[Theorem~1]{prc} for a detailed proof of this fact.
Our procedure also applies when the choice space is infinite.
In an independent working paper, \citet{petri2023ordered} develops a related
representation result that allows the underlying set of alternatives to be infinite.}
Consequently, the distribution induced by $r$ is the unique ordered representation
of $\rho$. Next, we show that $\{c^i\}_{i=1}^k\subseteq\mu$.

\begin{lemma}[Pairwise Replication]
\label{midlemma}
Let $\mu$ be a choice model such that    $\langle \mu, \vartriangleright\rangle$   is  a  lattice.\@ Let $c\in \C$ be a choice type.\@ If, for every pair $S,S'\in \X$, there exists a choice type $c^*\in \mu$ such that $\displaystyle{c^*(S)=c(S)}$ and $c^*(S')=c(S')$, then $c\in \mu$.                             
\end{lemma}
\begin{proof} 
%First, we make the following general observation that will be used recursively to obtain the result. 
The result is obtained by applying the following observation inductively.
Consider any   $\mathbb{S} \subset \X$  containing at least three menus. Let  $c_1, c_2, c_3\in \mu$ be such that for each $i\in \{1,2,3\}$,   there exists at most one $S_i\in \mathbb{S}$ with $c_i(S_i)\neq c(S_i)$.\@
%, and for each  $S\in \mathbb{S}\setminus \{S_i\}$,  $c_i(S)= c(S)$.\@ 
Suppose that for each distinct $i,j\in \{1,2,3\}$, if   such $S_i$ and $S_j$ exist, then $S_i\neq S_j$.\@ Now, for each $S\in \mathbb{S}$, the choice type $(c_1\wedge c_2)\vee  (c_1\wedge c_3) \vee(c_2\wedge c_3)\in \mu$ chooses $c(S)$.\@ To see this, let $S\in \mathbb{S}$, and note that there exist at least two $i,j\in \{1,2,3\}$ such that $c_i(S)=c_j(S)=c(S)$.
Assume without loss of generality that $i=1$ and $j=2$.\@ Now, if $c(S)\geq_S c_3(S)$, then  we get  $c(S)\vee c_3(S)\vee c_3(S)=c(S)$; if $c_3(S)>_S c(S)$, then we get $c(S)\vee c(S)\vee c(S)=c(S)$.\footnote{In this step, we establish that for any three menus, there is  a choice type  replicating $c$ across them. For the second step, for any four   menus,  four choice types\textendash replicating $c$ across each three of the four menus\textendash are similarly grouped   into sets of three  to replicate $c$  across the four menus.}  
\end{proof}

\begin{proof}[Proof of Theorem \ref{progressiveifflattice}] 
We proved the \textit{only if} part. For the \textit{if} part,  let      $c^r$ be a choice type that is assigned positive probability in the uniform decomposition  procedure.\@ We  show that     $c^r\in \mu$ by using Lemma \ref{midlemma}.\@  To see this, let   $S,S'\in \X$ such that $x=c^r(S)$ and $x'= c^r(S')$.\@ We will show that there exists $c^*\in \mu$ such that both $c^*(S)=x$ and $c^*(S')=x'$.\@ %Then, it will directly follow from Lemma \ref{midlemma} that $c^r\in \mu$.\@ 

First,  as demonstrated in Figure \ref{intervals}, we have $(1-l_{Sx})+ u_{S'x'}>1$.\@ Thinking probabilistically, this means that making a choice that is weakly lower than $x$ in $S$ and weakly higher than $x'$ in $S'$ are not  mutually  exclusive events.\@
Since  $\rho\in \Delta(\mu)$, it  follows that there exists $c_1\in \mu$ such that  $c_1(S)\leq_S x$ and $c_1(S')\geq_{S'} x'$.\@ Symmetrically, since  $(1-l_{S'x'})+ u_{Sx}>1$, there exists $c_2\in \mu$ such that  $c_2(S)\geq_S x$ and $c_2(S')\leq_{S'} x'$.\@ 
%\footnote{The argument in here is in the vein of \textit{probabilistic method} due to Paul Erd\H os. For a direct yet longer argument one can proced via contadiction.}
% can further clarify the claim for the reader. 
%?Next, let $\displaystyle{c_x=\bigvee_{\{c\in \rho^+ :\  x\geq_S c(S)\}} %c$ and  $\displaystyle{c_{x'}=\bigvee_{\{c\in \rho^+ : \ x'\geq_{S'} c(S')\}} %c$

Next,  consider the set $\{c\in \mu :  x\geq_S c(S)\}$ and let  $c_x$ be its join.\@ Similarly,   consider $\displaystyle{\{c\in \mu :  x'\geq_{S'} c(S')\}}$ and let  $c_{x'}$ be its join.\@ 
Note that  $c_1$ is an element  of the former set, while  $c_2$ is an element  of the latter one. Since 
$\rho\in\Delta(\mu)$ and $c^r$ is assigned positive probability in the uniform decomposition  procedure,    $\mu$ must contain a choice type  choosing $x$ from $S$  and a choice type  choosing $x'$ from $S'$.\@ Since  $\langle \mu, \vartriangleright\rangle$   is a lattice, it follows that   $\displaystyle{c_x(S)=x}$ and $\displaystyle{c_{x'}(S')=x'}$.\@ Now, define $c^*=c_x\wedge c_{x'}$.\@ Then,   $c^*(S)=x$, since    $c_x(S)=x$ and    $\displaystyle{c_{x'}(S)\geq_S c_2(S)\geq_S x}$.\@ Similarly,   $c^*(S')=x'$, since     $c_{x'}(S')=x'$ and    $\displaystyle{c_{x}(S')\geq_{S'} c_1(S')\geq_{S'} x'}$. Finally,  $c^*\in \mu$ since  $\langle \mu, \vartriangleright\rangle$   is a lattice containing $c_x$ and $c_{x'}$.
 \end{proof}

The uniform decomposition procedure can be implemented sequentially
using unnormalized residual choice probabilities, initialized at
$\rho$. At each step, the next choice type selects the highest-ranked
alternative with positive residual probability from each menu. Its
weight is the minimum, across menus, of the residual probabilities
of these selected alternatives. Subtracting this weight from the
residual probability of the selected alternative in each menu
reduces at least one positive residual probability to zero. Thus,
if the observed data contain $M$ menu--alternative pairs with
positive choice probabilities, the procedure terminates in at most
$M$ steps and yields an ordered representation supported on at most
$M$ choice types. The representation can therefore be recovered
in time polynomial in the size of the observed choice data,
without enumerating the potentially much larger space of all choice
types.

\section{Examples and discussion}
\label{chain}
%In Section \ref{latticedemo}, we present the choice types lattice for %is the example in Section \ref{latticedemo} demonstrates that probabilistic choice %model does not form a lattice even when there

\subsection{Rational choice and chain lattices}
\label{Section3.1}
%\textbf{Rational choice and chain lattices:}
We first observe that the rational choice model need not be
self-ordered.\@ To see this, let $X=\{a,b,c\}$ and
$\X=\{X,\{a,b\},\{a,c\},\{b,c\}\}$, and suppose that each primitive
ordering $>_S$ is obtained by restricting $a>b>c$ to menu $S$.\@
Figure~\ref{lattice} depicts the resulting lattice of all choice
types. Each four-letter string records, in order, the alternatives
chosen from $X$, $\{a,b\}$, $\{a,c\}$, and $\{b,c\}$.\@ The rational
choice types, shown in red, do not form a sublattice: some of their
joins and meets are nonrational types, shown in green.
\begin{figure}[h]
\label{monster}
\begin{center}

        \scalebox{0.45}{

 \begin{tikzpicture}

        %8th layer
        
        \node (l81) at (0.5,0) {$\boxed{\mathbf{aaab}}$};
        \node(81) at (0.5,-.5) {};
        \draw [fill = red](81) circle (4pt);

        %9th layer

        \node (l91) at (-4.5,-2.5) {$\boxed{\mathbf{baab}}$};
        \node(91) at (-4,-3) {};
        \draw [fill = green](91) circle (4pt);

        \node (l92) at (-1.5,-2.5) {$\boxed{\mathbf{abab}}$};
        \node(92) at (-1,-3) {};
        \draw [fill = black](92) circle (4pt);

        \node (l93) at (2.5,-2.5) {$\boxed{\mathbf{aacb}}$};
        \node(93) at (2,-3) {};
        \draw [fill = black](93) circle (4pt);
 
        \node (l94) at (5.5,-2.5) {$\boxed{\mathbf{aaac}}$};
        \node(94) at (5,-3) {};
        \draw [fill = red](94) circle (4pt);

%%%%%%%%%%%%%%%%%%%%%%%%%%        

        %10th layer
        \node (l107) at (-8.5,-5) {$\boxed{\mathbf{caab}}$};
        \node(107) at (-8,-5.5) {};
        \draw [fill = black](107) circle (4pt);

        \node (l100) at (-5.5,-5) {$\boxed{\mathbf{bbab}}$};
        \node(100) at (-5,-5.5) {};
        \draw [fill = red](100) circle (4pt);

        \node (l101) at (-2.5,-5) {$\boxed{\mathbf{bacb}}$};
        \node(101) at (-2,-5.5) {};
        \draw [fill = green](101) circle (4pt);

        \node (l102) at (1,-5) {$\boxed{\mathbf{baac}}$};
        \node(102) at (1,-5.5) {};
        \draw [fill = black](102) circle (4pt);

        \node (l103) at (4.5,-5) {$\boxed{\mathbf{abcb}}$};
        \node(103) at (4,-5.5) {};
        \draw [fill = black](103) circle (4pt);

        \node (l106) at (7.5,-5) {$\boxed{\mathbf{aacc}}$};
        \node(106) at (7,-5.5) {};
        \draw [fill = black](106) circle (4pt);

        \node (l104) at (10.5,-5) {$\boxed{\mathbf{abac}}$};
        \node(104) at (10,-5.5) {};
        \draw [fill = black](104) circle (4pt);

        %11th layer
        \node (l110) at (-8.6,-8) {$\boxed{\mathbf{caac}}$};
        \node(110) at (-8,-8.5) {};
        \draw [fill = black](110) circle (4pt);

        \node (l111) at (-5.7,-8) {$\boxed{\mathbf{bbcb}}$};
        \node(111) at (-5,-8.5) {};
        \draw [fill = red](111) circle (4pt);
        
        \node (l112) at (-2.5,-8) {$\boxed{\mathbf{cbab}}$};
        \node(112) at (-2,-8.5) {};
        \draw [fill = black](112) circle (4pt);

        \node (l113) at (1,-8) {$\boxed{\mathbf{bbac}}$};
        \node(113) at (1,-8.5) {};
        \draw [fill = green](113) circle (4pt);

        \node (l114) at (4.5,-8) {$\boxed{\mathbf{cacb}}$};
        \node(114) at (4,-8.5) {};
        \draw [fill = black](114) circle (4pt);
        
        \node (l115) at (7.5,-8) {$\boxed{\mathbf{bacc}}$};
        \node(115) at (7,-8.5) {};
        \draw [fill = black](115) circle (4pt);

        \node (l116) at (10.5,-8) {$\boxed{\mathbf{abcc}}$};
        \node(116) at (10,-8.5) {};
        \draw [fill = black](116) circle (4pt);

        %12th layer
        
        \node (l121) at (-4.5,-11.5) {$\boxed{\mathbf{cacc}}$};
        \node(121) at (-4,-11) {};
        \draw [fill = red](121) circle (4pt);

        \node (l123) at (-1.5,-11.5) {$\boxed{\mathbf{cbac}}$};
        \node(123) at (-1,-11) {};
        \draw [fill = black](123) circle (4pt);

        \node (l124) at (2.5,-11.5) {$\boxed{\mathbf{cbcb}}$};
        \node(124) at (2,-11) {};
        \draw [fill = black](124) circle (4pt);

        \node (l122) at (5.5,-11.5) {$\boxed{\mathbf{bbcc}}$};
        \node(122) at (5,-11) {};
        \draw [fill = green](122) circle (4pt);

        %13th layer

        \node (l131) at (0.5,-14) {$\boxed{\mathbf{cbcc}}$};
        \node(131) at (0.5,-13.5) {};
        \draw [fill = red](131) circle (4pt);
        
        %%%%%%%%%%%%%%%%%%%%%%%%%%%%%%          
         %%%%%%%%%%%%%%%%%%%%%%%%%%%%%%    

        %lines 8th on

        \draw [dashed, thick] (81) to [] (91); 
        
        \draw [dashed, thick] (81) to [] (92);

        \draw [dashed, thick] (81) to [] (93);
        
        \draw [dashed, thick] (81) to [] (94);

        \draw [dashed, thick] (91) to [] (107); 
        
        \draw [dashed, thick] (91) to [] (100);

        \draw [dashed, thick] (91) to [] (101);
        
        \draw [dashed, thick] (91) to [] (102);

\draw [dashed, thick] (92) to [] (100); 
        \draw [dashed, thick] (92) to [] (103);        
        \draw [dashed, thick] (92) to [] (104);

\draw [dashed, thick] (93) to [] (101); 
        \draw [dashed, thick] (93) to [] (103);        
        \draw [dashed, thick] (93) to [] (106);

\draw [dashed, thick] (94) to [] (102); 
        \draw [dashed, thick] (94) to [] (104);        
        \draw [dashed, thick] (94) to [] (106);

\draw [dashed, thick]  (121)to [] (131); 
        
\draw [dashed, thick]  (123)to [] (131); 

\draw [dashed, thick]  (122)to [] (131);

  \draw [dashed, thick] (124)to [] (131);

\draw [dashed, thick] (121) to [] (110); 

\draw [dashed, thick] (121) to [] (115); 

\draw [dashed, thick] (121) to [] (114);

\draw [dashed, thick] (122) to [] (116); 

\draw [dashed, thick] (122) to [] (115); 

\draw [dashed, thick] (122) to [] (113); 
\draw [dashed, thick] (122) to [] (111);

\draw [dashed, thick] (123) to [] (113); 

\draw [dashed, thick] (123) to [] (110); 

\draw [dashed, thick] (123) to [] (112);

\draw [dashed, thick] (124) to [] (112); 

\draw [dashed, thick] (124) to [] (114); 

\draw [dashed, thick] (124) to [] (111); 

%%%%%%%%%%%%%%%%
\draw [dashed, thick] (107) to [] (110); 

\draw [dashed, thick] (107) to [] (114); 

\draw [dashed, thick] (107) to [] (112);

\draw [dashed, thick] (100) to [] (113); 

\draw [dashed, thick] (100) to [] (111); 

\draw [dashed, thick] (100) to [] (112);

\draw [dashed, thick] (101) to [] (114); 

\draw [dashed, thick] (101) to [] (111); 

\draw [dashed, thick] (101) to [] (115);

\draw [dashed, thick] (102) to [] (110); 

\draw [dashed, thick] (102) to [] (113); 

\draw [dashed, thick] (102) to [] (115);

\draw [dashed, thick] (103) to [] (111); 

\draw [dashed, thick] (103) to [] (116);

\draw [dashed, thick] (104) to [] (113); 

\draw [dashed, thick] (104) to [] (116); 

\draw [dashed, thick] (106) to [] (115); 

\draw [dashed, thick] (106) to [] (116);

        \end{tikzpicture}
}

\end{center}
\caption{The lattice of choice types, where each node lists the alternatives chosen from the menus $\{a,b,c\}$, $\{a,b\}$, $\{a,c\}$, and $\{b,c\}$, respectively, and dotted lines correspond to the comparison relation  $\vartriangleright$ (obtained from $>$) among choice types. The rational choice types are shown in red and their joins and meets in green; for example, $[baab]$ is the join of $[bbab]$ and $[cacc]$.}
\label{lattice}
\end{figure}
%Choice types lattice for $X=\{a,b,c\}$, $\X$ is all menus, and %the unique primitive ordering is $a>b>c$.
%%%%%%%%%%%%%%%%%%%%%%%%%%%%%%%%%%%%%%%%%%%%%%%%%%%%%%%%%%%%%%%%

%\vspace{.7cm}
%\

The lattice characterization provides a guide for restricting or
extending the rational choice model to make it self-ordered.\@ A
particularly simple example is a \textit{chain lattice}: a collection
of choice types $\{c_i\}_{i=1}^k$ that are totally ordered by
$\vartriangleright$. After relabeling if necessary, they satisfy
$
c_1\vartriangleright c_2\vartriangleright\cdots
\vartriangleright c_k.
$
Suppose now that each primitive ordering $>_S$ is obtained by
restricting a common ordering $>_X$ to menu $S$.\@ Collections of
preferences satisfying the single-crossing property defined by
\citet{scrum} induce chain lattices of rational choice types on any
choice domain.
The converse---that a given collection of rationalizing preferences is
single-crossing whenever its choice types are ordered---holds when the
choice domain contains every binary menu.\@ With restricted observations,
it may fail because comparisons on unobserved binary menus need not be
revealed.
%the preferences rationalizing the choice types in $\mu$ 

%See also  \cite{curello2019preference} who examine when a primitive ordering %over alternatives allows for lattice structure of single-crossing dominance %over preferences and its applications.
%  %and provide charaterizing conditions on partial orders of alternatives %for minimum upper bounds to exist for every set of preferences e 
%proving characterisation, existence and uniqueness results for minimum
%upper bounds of arbitrary sets of preferences.
%} 

\subsection{Multiple behavioral characteristics} 
%\vspace*{-.21cm}
%\noindent \textbf{Beyond chain lattices:} 
\label{Section3}
A probabilistic choice function $\rho$ is \emph{ordered} if it admits
a representation supported on $\vartriangleright$-comparable choice
types, without requiring those types to belong to any prespecified
choice model. By contrast, $\langle\mu,\vartriangleright\rangle$ is a chain
lattice if all choice types in $\mu$ are pairwise
$\vartriangleright$-comparable. \citet{scrum} provide economic
examples of rational choice types that form chain lattices, whereas
\citet{prc} present nonrational choice models that capture a single
behavioral characteristic and likewise form chain lattices.
Theorem~\ref{progressiveifflattice} shows, however, that self-ordered
choice models need not be chain lattices. Such models may contain
incomparable choice types and can therefore accommodate multiple
behavioral characteristics indexed by distinct parameters. The
following example illustrates the economic value of this additional
flexibility.
%In our next example, we present a choice model in which  agents   deviate %from their
%accurate preferences  when they choose from larger menus.

%\footnote{See  also   \cite{curello2019preference} who characterize when %a common primitive ordering over alternatives allows for preferences  form %a lattice according to  single-crossing dominance, and provide several applications.

%\footnote{See \cite{chernev2015choice} for a recent meta-analysis about choice overload phenomena.} %demonstrating the generality and economic relevance of self-ordered %choice models.

%\newpage

%\begin{example}{(Choice overload)} \normalfont  Let $\mathcal{P}$   be  %a  set of faulty preferences  that are single-crossing with respect to the %true preference   $>$.\@ Then, a choice type $c\in  \mu$ if (1) for %each menu $S$, the alternative $c(S)$ is the $\succ_S$-maximal one %in $S$ for some $\succ_S\in \mathcal{P}$,  and  (2) $\succ_S$ is  aligned % with $>$ as much as  $\succ_{S'}$ is aligned  with $>$, whenever $S$ is %a subset of $S'$. % (2) $\succ_S$ is as much   aligned  with $>$ as   $\succ_{S'}$, %whenever $S$ is a subset of $S'$.\@ 
%Note that    $\mu$ is self-ordered with respect to 
%the comparison relation  obtained from $>$, since for each $c^i,c^j\in \mu$, their join and meet are the choice types described by  maximization of the preferences   $max(\{\succ^i_S,\succ^j_S\},\geq)$  and $\displaystyle{min(\{\succ^i_S,\succ^j_S\},\geq)}$. %\end{example}

\begin{example}[Similarity and Choice under Risk]
\label{similarity}
\normalfont
Following the choice-by-similarity procedure introduced by
\citet{similarity}, consider binary lotteries of the form
$a=(m_a,p_a)$. Lottery $a$ pays $m_a$ with probability $p_a$ and zero
otherwise. Assume that $m_a,p_a\in[0,1]$. Consider a population of
agents who choose from binary menus of these lotteries.
\footnote{To keep the choice domain finite, one may assume that
monetary outcomes and probabilities vary in fixed positive
increments.}
The primitive ordering ranks lotteries according to their expected
monetary payoffs: it ranks $a$ above $b$ whenever
$m_ap_a>m_bp_b$, with ties resolved according to a fixed rule.

A decision maker may disregard a relatively small difference in one
feature when the lotteries differ substantially in the other. For
each choice type $i$, we represent this behavior using two weakly
increasing threshold functions, $g_p^i$ and $g_m^i$, satisfying
$g_p^i(d)\geq d$ and $g_m^i(d)\geq d$ for every $d\in[0,1]$.

Consider two lotteries $a$ and $b$. Suppose that $a$ offers the larger
prize, so that $m_a>m_b$, while $b$ offers the greater probability of
winning, so that $p_b>p_a$. Let $d_m=m_a-m_b$ denote the prize
difference and $d_p=p_b-p_a$ the difference in winning probabilities.
Type $i$ chooses $b$ if $b$'s probability advantage is sufficiently
large relative to $a$'s prize advantage, that is, if
$d_p>g_p^i(d_m)$. Conversely, type $i$ chooses $a$ if its prize
advantage is sufficiently large relative to $b$'s probability
advantage, that is, if $d_m>g_m^i(d_p)$. If neither inequality holds,
the type chooses the lottery ranked higher by the reference ordering
$>$, namely, the one with the higher expected monetary payoff, using
the fixed tie-breaking rule when expected payoffs are equal. The same
expected-payoff ranking, including its tie-breaking rule, applies when
one lottery dominates the other in both features or when the lotteries
have the same prize or the same winning probability.

Thus, $g_p^i(d_m)$ is the threshold that the probability advantage
must exceed to become decisive when the prize difference is $d_m$.
Conversely, $g_m^i(d_p)$ is the threshold that the prize advantage
must exceed to become decisive when the probability difference is
$d_p$. The restrictions $g_p^i(d)\geq d$ and $g_m^i(d)\geq d$ ensure
that the two rules are mutually exclusive: the first can apply only
if $d_p>d_m$, whereas the second can apply only if $d_m>d_p$.

This procedure provides an explanation for the Allais paradox
\citep[cf.][]{similarity}. Taking the expected-payoff ranking as the
primitive ordering, the analyst can use the model to describe
population heterogeneity arising from different degrees of bounded
rationality.\@ The question is whether the resulting
similarity-based choice model is self-ordered---that is, whether every
\rcf\ it explains admits an ordered representation within the model.
We show next that the answer is affirmative.

The collection of similarity-based choice types forms a lattice. To
establish closure under joins, consider any two types $i$ and $j$.
Define the type $i\vee j$ by
$g_p^{i\vee j}(d)=\max\{g_p^i(d),g_p^j(d)\}$ and
$g_m^{i\vee j}(d)=\max\{g_m^i(d),g_m^j(d)\}$.
An advantage exceeds the maximum threshold only if it exceeds the
thresholds of both types. Thus, a given feature is decisive for
$i\vee j$ only when it is decisive for both $i$ and $j$. In all other
cases, $i\vee j$ follows the expected-payoff ranking and therefore
selects the $>$-higher of the alternatives chosen by the two types.
Hence, on every binary menu, $i\vee j$ selects the $>$-higher
alternative among the choices of $i$ and $j$, so it is their join.

Closure under meets follows similarly. Define the type $i\wedge j$ by
$g_p^{i\wedge j}(d)=\min\{g_p^i(d),g_p^j(d)\}$ and
$g_m^{i\wedge j}(d)=\min\{g_m^i(d),g_m^j(d)\}$.
An advantage exceeds the minimum threshold whenever it exceeds the
threshold of at least one of the two types. Moreover, a probability
advantage can be decisive only when $d_p>d_m$, whereas a prize
advantage can be decisive only when $d_m>d_p$. Thus, the two types
cannot regard different features as decisive for the same pair of
lotteries. The type $i\wedge j$ can therefore follow any feature
regarded as decisive by either type without creating a conflict. On
every binary menu, it selects the $>$-lower alternative among the
choices of $i$ and $j$, so it is their meet.

Pointwise maxima and minima preserve both weak monotonicity and the
restrictions $g_p^i(d)\geq d$ and $g_m^i(d)\geq d$. Hence, the
constructed joins and meets remain similarity-based choice types.
The similarity-based choice model therefore forms a lattice and, by
Theorem~\ref{progressiveifflattice}, is self-ordered.
\end{example}

\subsection{Examples in related economic environments}
\label{Section3.3}
The lattice characterization also connects our framework to several
familiar economic environments. The next three examples consider
monotone demand, Walrasian equilibrium prices, and stable matchings.
In each case, the relevant deterministic objects admit a natural
partial order, and established results---or a direct argument---show
that the admissible objects are closed under the corresponding joins
and meets. By interpreting these objects as types and aggregate
observations as mixtures over them, we can apply
Theorem~\ref{progressiveifflattice} to obtain an ordered
representation of the aggregate data.

\begin{example}[Monotone comparative statics]
\label{mcs}
\normalfont
Let $X$ be a finite set of pension plans, linearly ordered by simplicity
under $>$. Agents in the population face menus $S\subset X$ from which
they select a plan. A menu $S$ \textit{dominates} a menu $T$ in the
\textit{strong set order} if, for every $x\in S$ and $y\in T$, the
simpler of the two plans belongs to $S$, while the more complex plan
belongs to $T$. This is the standard strong set order introduced by
\citet{Veinott1989} and widely used in monotone comparative statics
\citep{MilgromShannon1994,Topkis1998}.

We call a choice type $c$ \textit{monotone} if it is nondecreasing with
respect to the strong set order: for every pair of menus $S,T\in\X$, if
$S$ dominates $T$, then $c(S)\geq c(T)$ according to the order induced
by $>$.%
\footnote{In particular, if an agent maximizes a utility function with
\textit{increasing differences} \citep{topkis1978minimizing}, the
resulting demand function (choice type) is nondecreasing with respect to
the strong set order. Conversely, any such demand function can be
rationalized by a menu-dependent utility function with increasing
differences.}

Suppose an analyst observes choice probabilities over pension
plans from each menu and that these probabilities admit a
representation using monotone choice types. The question is
whether they also admit an ordered representation using only
such types. Theorem~\ref{progressiveifflattice} answers this
question affirmatively, and the ordered representation is unique.
Indeed, the pointwise join and meet of any two monotone
choice types select, respectively, the simpler and the more complex of
the two plans chosen from each menu. Both resulting choice types remain
monotone. Hence, the set of monotone choice types forms a lattice and,
by Theorem~\ref{progressiveifflattice}, is self-ordered.
This result turns an otherwise underidentified mixture of heterogeneous
demand rules into a unique ordered representation. It thereby allows
aggregate choice probabilities to reveal how the population is
distributed across simpler and more complex behavioral responses.
\end{example}

        %In particular, we consider a competitive equilibrium setting in which types correspond to equilibrium price vectors. The analyst observes only the frequencies with which different equilibrium prices arise across individual markets. The objective is to represent these probabilistic observations in an ordered manner—so that equilibrium price vectors can be compared from lower to higher—for purposes such as comparative statics, policy evaluation, or to provide an informative summary of the data.

\begin{example}[Competitive equilibrium prices]
\label{walraslattice}
\normalfont
Consider a discretized economy with finitely many indivisible goods.
Consumers have preferences over bundles of goods and money that are
quasilinear in money. The price of each good is restricted to a
sufficiently fine finite grid
$\{p_0,p_0+\epsilon,\dots,p_0+K\epsilon\}\subset\mathbb{R}_+$, where
$\epsilon$ denotes the smallest unit of currency. Let $X$ denote the
set of all price vectors on this grid. We order these vectors
coordinatewise using the usual order on real numbers. Thus, one price
vector is higher than another if it assigns a weakly higher price to
every good and a strictly higher price to at least one good.

%\

Suppose an analyst observes, across different economic environments,
only the frequency with which each equilibrium price arises for each
good. We interpret each equilibrium price vector as a choice type. The
analyst seeks to represent these marginal price distributions as a
mixture of equilibrium price vectors whose support is ordered
coordinatewise from lower to higher prices. Such a representation
provides an informative summary of equilibrium heterogeneity and can
facilitate comparative-static and policy analysis.

%\vspace{-.1cm}

By Theorem~\ref{progressiveifflattice}, the desired unique ordered
representation is guaranteed if and only if the set of Walrasian
equilibrium price vectors forms a lattice. \citet{GulStacchetti1999}
show that this lattice property holds when consumers' preferences
satisfy the \textit{single improvement property}.%
\footnote{In the terminology of Gul and Stacchetti, a utility function
satisfies the \textit{single improvement property} if, whenever a
bundle is not optimal at a given price vector, another bundle yields
strictly higher net utility and differs from the original bundle by
adding one good, removing one good, or doing both.}
Under the coordinatewise order, the maximum and minimum of any two
Walrasian equilibrium price vectors are therefore also equilibrium
price vectors. For the discretized setting considered here,
\citet{murota2016discrete}, building on \citet{ausubel2006efficient},
establishes the corresponding result for integer-valued prices using
discrete convex analysis.%
\footnote{See Theorem 8.3 in \citet{murota2016discrete}, which shows
that the equilibrium price vectors form a bounded
$L^\natural$-convex set. See also the corollary to Proposition 1 in
\citet{ausubel2006efficient}.}
Hence, by Theorem~\ref{progressiveifflattice}, the observed price
distributions admit the desired unique ordered representation from
lower to higher equilibrium price vectors.
This is useful because marginal price frequencies alone generally do
not identify how equilibrium prices co-move across goods. The result
selects a unique ordered joint representation of this variation,
thereby supporting more disciplined comparative-static and policy
analyses.

%Minimizers of a Lyupunov submodular function
\end{example}

%\vspace*{-.3cm}

\begin{example}[Stable matchings]
\label{stablematchinglattice}
\normalfont
Consider a one-to-one marriage market with equal numbers of men and
women, each of whom has strict preferences over members of the other
side. A matching is \textit{stable} if there is no man--woman pair who
both prefer each other to their assigned partners. To fit this
environment into our framework, we interpret each stable matching as a
choice type. We order stable matchings by the men-wise dominance
relation: one stable matching is higher than another if every man
weakly prefers his partner under the former matching to his partner
under the latter. That is, each man either has the same partner under both matchings or
strictly prefers his partner under the former matching to his partner under the latter.

%\vspace*{-.2cm}
Consider an analyst who observes only a probabilistic assignment, which
specifies, for each man--woman pair, the probability that they are
matched. A probabilistic assignment is \textit{ex-post stable} if it
can be generated by a probability distribution over deterministic
stable matchings. Since stable matchings form a lattice under the men-wise
dominance relation
\citep{knuth1976marriages,rothsotomayor1990},
Theorem~\ref{progressiveifflattice} implies that every ex-post stable
probabilistic assignment admits a unique ordered decomposition into
stable matchings whose support is ranked from better to worse for men.%
\footnote{The existence part recovers the \textit{well-ordered stable
decomposition} established by \citet{dogan2015efficient}. See
\citet{juarez2022lattice} for an extension to many-to-many matching
markets.}
This may be important because pairwise matching probabilities alone generally
do not identify the distribution of stable matchings that generated
them. The unique ordered decomposition provides a disciplined account
of this latent heterogeneity and supports comparisons across
probabilistic assignments of the kind studied by
\citet{juarez2022lattice}.
\end{example}

\vspace{-.5cm}
\section{Minimal self-ordered extension of rational choice}
\label{minimalext}
In Section~\ref{Section3.1}, we showed that the rational choice model
fails to be self-ordered. Specifically, there exist RUMs whose ordered
representations require positive weight on nonrational choice types.
This raises two central questions: What structure must these additional
nonrational choice types have, and do they admit an economic
interpretation?

To answer these questions, we apply the recipe provided by
Theorem~\ref{progressiveifflattice} to identify the minimal self-ordered
extension of the rational choice model. An extension is
\textit{minimal} if it adds nonrational choice types as sparingly as
possible: it is self-ordered and contains all rational choice types, but
no proper subset containing all rational choice types is self-ordered.%
\footnote{Theorem~\ref{progressiveifflattice} implies that the minimal
extension of any choice model is unique.}
Throughout this section, we assume that there is a single primitive
ordering \(>\) that ranks alternatives according to their ``accurate
values,'' and that every subset of the alternative set containing at
least two alternatives is observable as a menu.

We now characterize the minimal self-ordered extension of the rational
choice model in terms of two choice axioms. The weak axiom of revealed
preference (WARP) provides a natural point of departure. In the
single-valued setting considered here, WARP implies the following
contraction-consistency condition: an alternative chosen from a menu
must remain chosen from every smaller menu in which it remains
available. Thus, if $c(S)=x$ and $x\in T\subseteq S$, then $c(T)=x$.

When alternatives are endowed with a primitive ordering, our axioms can
be viewed as refinements of this consistency requirement. They do not
require the selected alternative to remain unchanged following a menu
contraction or expansion. Instead, they permit choice to change while
restricting the direction of that change according to the primitive
ordering. In particular, the quality of the selected alternative cannot
decline when an inferior alternative is removed or a superior
alternative is added.

\noindent \textbf{Inferior-Removal Monotonicity (IRM).}
Removing an alternative ranked below the selected alternative cannot
result in a lower-ranked choice. Formally, a choice type $c$ satisfies
IRM if, for every menu $S\in\X$ with $|S|\geq 3$ and every
$x\in S$,
\[
\text{if } c(S)>x,\text{ then }
c(S\setminus\{x\})\geq c(S).
\]

\noindent \textbf{Superior-Addition Monotonicity (SAM).}
Adding an alternative ranked above the choice from the expanded menu
cannot result in a choice ranked below the choice from the original
menu.%
\footnote{The \emph{independence from preferred alternatives} axiom of
\citet*{masatlioglu2020willpower} similarly requires the choice to
remain unchanged when unchosen higher-ranked alternatives are removed.}
Formally, a choice type $c$ satisfies SAM if, for every menu $S\in\X$
with $|S|\geq 3$ and every $x\in S$,
\[
\text{if } x>c(S),\text{ then }
c(S)\geq c(S\setminus\{x\}).
\]

A choice type is rational if and only if it satisfies WARP. IRM and SAM
retain the underlying intuition of choice consistency while replacing
exact choice invariance with restrictions on how choice may respond to
menu changes.

More broadly, expanding a menu can change the selected alternative in
either direction: it may result in a lower-ranked choice, but it may also
make a higher-ranked alternative available. IRM and SAM discipline these
changes according to the rank of the added or removed alternative.
Together, these axioms allow the model to accommodate boundedly rational
choice types while preserving a disciplined relationship between menu
changes and the quality of the selected alternative. The following
result shows that they identify precisely the additional choice types
required for the minimal self-ordered extension of rational choice.

\begin{theorem}[Minimal self-ordered extension of rational choice]
\label{minimalextension}
A choice type belongs to the minimal self-ordered extension of the
rational choice model if and only if it satisfies IRM and SAM.
\end{theorem}

\begin{proof}
Please see Section~\ref{minimalproof} in the Appendix.
\end{proof}

Figure~\ref{rationalextension} illustrates this minimal extension for
the case of three alternatives. The proof of
Theorem~\ref{minimalextension} also demonstrates how
Theorem~\ref{progressiveifflattice} and Lemma~\ref{midlemma} can be
used to derive analogous characterizations for other choice models.

 %%%%%%%%%%%%%%%%%%%%%%%%%%%%%%%%%%%%%%%%%%%%%%%%
%%%%%%%%%%%%%%%%%%%%%%%%%%%%%%%%%%%%%%%%%%%%%%%%%%%%

\begin{figure}[h]

\begin{center}

        \scalebox{0.45}{

 \begin{tikzpicture}

        %8th layer
        
        \node (l81) at (0.5,0) {$\boxed{\mathbf{aaab}}$};
        \node(81) at (0.5,-.5) {};
        \draw [fill = red](81) circle (4pt);

        %9th layer

        \node (l91) at (-1.5,-2.5) {$\boxed{\mathbf{baab}}$};
        \node(91) at (-1,-3) {};
        \draw [fill = green](91) circle (4pt);

        \node (l94) at (2.5,-2.5) {$\boxed{\mathbf{aaac}}$};
        \node(94) at (2,-3) {};
        \draw [fill = red](94) circle (4pt);

%%%%%%%%%%%%%%%%%%%%%%%%%%        

        %10th layer

        \node (l101) at (-3.1,-5) {$\boxed{\mathbf{bacb}}$};
        \node(101) at (-2.6,-5.5) {};
        \draw [fill = green](101) circle (4pt);

        \node (l102) at (4.1,-5) {$\boxed{\mathbf{baac}}$};
        \node(102) at (3.6,-5.5) {};
        \draw [fill = orange](102) circle (4pt);

        \node (l100) at (.5,-5) {$\boxed{\mathbf{bbab}}$};
        \node(100) at (.5,-5.5) {};
        \draw [fill = red](100) circle (4pt);

        %11th layer

        \node (l111) at (.5,-7.5) {$\boxed{\mathbf{bbcb}}$};
        \node(111) at (.5,-8) {};
        \draw [fill = red ](111) circle (4pt);

       \node (l115) at (-3.1,-7.5) {$\boxed{\mathbf{bacc}}$};
        \node(115) at (-2.6,-8) {};
        \draw [fill = orange](115) circle (4pt);

        \node (l113) at (4.1,-7.5) {$\boxed{\mathbf{bbac}}$};
        \node(113) at (3.6,-8) {};
        \draw [fill = green](113) circle (4pt);

        %12th layer
        
        \node (l121) at (-1.5,-10) {$\boxed{\mathbf{cacc}}$};
        \node(121) at (-1,-10.5) {};
        \draw [fill = red](121) circle (4pt);

        \node (l122) at (2.5,-10) {$\boxed{\mathbf{bbcc}}$};
        \node(122) at (2,-10.5) {};
        \draw [fill = green](122) circle (4pt);

        %13th layer

        \node (l131) at (0.5,-13.5) {$\boxed{\mathbf{cbcc}}$};
        \node(131) at (0.5,-13) {};
        \draw [fill = red](131) circle (4pt);
        
        %%%%%%%%%%%%%%%%%%%%%%%%%%%%%%          
         %%%%%%%%%%%%%%%%%%%%%%%%%%%%%%    

        %lines 8th on

        \draw [dashed, thick] (81) to [] (91);

        \draw [dashed, thick] (81) to [] (94);

        \draw [dashed, thick] (91) to [] (100);

        \draw [dashed, thick] (91) to [] (101);
        
        \draw [dashed, thick] (91) to [] (102);

\draw [dashed, thick] (94) to [] (102);

\draw [dashed, thick]  (121)to [] (131);

\draw [dashed, thick]  (122)to [] (131);

\draw [dashed, thick] (122) to [] (115); 

\draw [dashed, thick] (122) to [] (113); 
\draw [dashed, thick] (122) to [] (111);

%\draw [dashed, thick] (124) to [] (111); 

%%%%%%%%%%%%%%%%

\draw [dashed, thick] (100) to [] (113); 

\draw [dashed, thick] (100) to [] (111);

\draw [dashed, thick] (115) to [] (121);

\draw [dashed, thick] (101) to [] (111); 

\draw [dashed, thick] (101) to [] (115);

\draw [dashed, thick] (102) to [] (113); 

\draw [dashed, thick] (102) to [] (115); 

\end{tikzpicture}
}

\end{center}
\caption{The minimal self-ordered extension of the rational choice
model under $\vartriangleright$, where $X=\{a,b,c\}$,
$\X=\{X,\{a,b\},\{a,c\},\{b,c\}\}$, and each array represents a
choice type by listing its choices. Rational choice types are shown in
red, their joins and meets in green, and the additional choice types
obtained as joins or meets of the preceding types in orange.}
\label{rationalextension}
\end{figure}
%
%%%%%%%%%%%%%%%%%%%%%%%%%%%%%%%%%%%%%%%%%%%%%%%%%%%%%%%%%%%%%%%%

What choice patterns can the model characterized by IRM and SAM
accommodate that the random utility model cannot? The following example
shows that IRM and SAM permit a disciplined form of context dependence
that can generate the attraction effect, which the random utility model
rules out. The attraction effect is a well-documented pattern in which
adding an asymmetrically dominated alternative increases the probability
of choosing the alternative that dominates it.%
\footnote{The attraction effect was first documented experimentally by
\citet{HuberPaynePuto1982} and \citet{HuberPuto1983}. Subsequent studies
have documented the effect in a wide variety of settings; for a review,
see \citet{RieskampBusemeyerMellers2006}.}

\begin{example}[Attraction effect]\normalfont
\label{attractionexample}

Let $X=\{a,b,c\}$ denote three products evaluated along two dimensions:
price and quality. Product $a$ is more attractive than $b$ in terms of
price, whereas $b$ is more attractive than $a$ in terms of quality.
Product $c$ is an asymmetrically dominated decoy: product $b$ dominates
it on both dimensions, whereas product $a$ does not. Specifically,
suppose that the two criteria rank the products as
$a\succ_{\mathrm{pr}}b\succ_{\mathrm{pr}}c$ and
$b\succ_{\mathrm{q}}c\succ_{\mathrm{q}}a$.
Let the primitive ordering $>$ coincide with the price ranking, so that
$a>b>c$. Then, consider the following three choice types and their population weights:

\begin{center}
\begin{tabular}{c|c|c|c|c|c}
 & Weight & $\{a,b,c\}$ & $\{a,b\}$ & $\{a,c\}$ & $\{b,c\}$ \\ \hline
$c_1$ & $1/3$ & $a$ & $a$ & $a$ & $b$ \\
$c_2$ & $1/2$ & $b$ & $b$ & $c$ & $b$ \\
$c_3$ & $1/6$ & $b$ & $a$ & $c$ & $b$
\end{tabular}
\end{center}

The first two choice types are rational: type $c_1$ maximizes the price
ranking, whereas type $c_2$ maximizes the quality ranking. Type $c_3$
exhibits the attraction effect. When only $a$ and $b$ are available, it
selects $a$. When the decoy $c$ is added, the dominance of $b$ over $c$
makes $b$ more attractive, leading the type to switch to $b$.

Let $\rho^*$ denote the $\rcf$ generated by these choice types with the
indicated population weights. From the menu $\{a,b\}$, types $c_1$ and
$c_3$ choose $a$, whereas type $c_2$ chooses $b$. Hence,
$\rho^*_{b}(\{a,b\})=1/2$. From the menu $\{a,b,c\}$, types $c_2$ and
$c_3$ choose $b$, whereas type $c_1$ chooses $a$. Thus,
$\rho^*_{b}(\{a,b,c\})=2/3$. Adding the dominated alternative $c$
therefore increases the probability of choosing $b$. Consequently, no
random utility model can generate this pattern.

The additional type $c_3$ is nonrational: it chooses $b$ from
$\{a,b,c\}$ but $a$ from the smaller menu $\{a,b\}$. Nevertheless, it
satisfies both IRM and SAM. The only nontrivial application of IRM
arises when $c$ is removed from $\{a,b,c\}$. Since
$c_3(\{a,b,c\})=b>c$ and $c_3(\{a,b\})=a>b$, this removal improves the
choice, as required by IRM. SAM also holds: removing the superior
alternative $a$ from $\{a,b,c\}$ leaves the choice unchanged.
\end{example}

The example highlights the behavioral content of the minimal extension.
It allows a dominated alternative to affect choice, but only in a
direction disciplined by the primitive ordering. Here, removing the
inferior decoy changes the choice from $b$ to the higher-ranked
alternative $a$. Thus, the model accommodates the context dependence
underlying the attraction effect while ruling out unrestricted
departures from rational choice.

How difficult is it to construct the minimal self-ordered extension?
This question is particularly relevant for applications. The extension
is obtained by closing the model under pointwise joins and meets.
This construction is related in spirit to classical lattice-completion
problems. For example, the Dedekind--MacNeille completion is unique,
and its construction for finite partially ordered sets is well
understood. Although its size may grow exponentially in the worst
case, it can be constructed incrementally, with each step taking
polynomial time.%
\footnote{Each step of the algorithm proposed by
\citet{ganter1998stepwise} runs in time polynomial in the size and width
of the partially ordered set. \citet{nourine1999fast} subsequently
provide a faster algorithm with an improved complexity bound.}

% Replace the existing Section 4.1 with the first block below.
% Replace the existing Appendix subsection on Theorem 3 and Corollary 1
% with the second block. The existing preamble and bibliography are used.

%%%%%%%%%%%%%%%%%%%%%%%%%%%%%%%%%%%%%%%%%%%%%%%%%%%%%%%%%%%%%%%%
% SECTION 4.1
%%%%%%%%%%%%%%%%%%%%%%%%%%%%%%%%%%%%%%%%%%%%%%%%%%%%%%%%%%%%%%%%

\subsection{Identification of the primitive ordering}

So far, we have assumed either that alternatives are naturally ordered
or that the analyst has specified the primitive ordering. We now study
how to identify a primitive ordering with respect to which a given
choice model is contained in the minimal self-ordered extension of the
rational choice model. The analysis takes the model's admissible
deterministic choice types as given, consistent with the analyst's
model-selection problem discussed in the introduction.

Formally, a primitive ordering $>$ is \textit{compatible with} a choice
model $\mu$ if $\mu$ is contained in the minimal self-ordered extension
of the rational choice model with respect to $>$. By
Theorem~\ref{minimalextension}, this is equivalent to requiring every
choice type in $\mu$ to satisfy IRM and SAM according to $>$. The choice
types in $\mu$ can then be interpreted as sampled choice behaviors from
a population whose members satisfy IRM and SAM according to $>$.

We first show how to infer from a given choice model that an alternative
lies ``between'' two other alternatives according to every primitive
ordering compatible with the model.
Removing an unchosen alternative can change the selected alternative.
Under IRM and SAM, such a switch reveals that the original choice lies
between the new choice and the removed alternative in the primitive
ordering.

Formally, let $\mu$ be a given choice model and let $x,y,z\in X$ be
three distinct alternatives. We say that $y$ is \textbf{revealed to be
between} $x$ and $z$---denoted by $\B{y}{x}{z}$---if, after possibly
interchanging $x$ and $z$, there exists a choice type $c\in\mu$ such
that $c(S)=y$ and $c(S\setminus\{z\})=x$ for some menu $S$.
For every primitive ordering $>$ compatible with $\mu$, if
$\B{y}{x}{z}$, then $x>y>z$ or $x<y<z$. We refer to
$\mathcal{B}_\mu$ as the \textbf{betweenness relation} associated with
the choice model $\mu$.

We next introduce a \textit{richness} condition on the betweenness relation
$\mathcal{B}_\mu$. Under this condition, we characterize when a
compatible primitive ordering exists and show that it is unique up
to reversal. We also show that the primitive ordering associated with
the minimal self-ordered extension of the rational choice model is
identified up to reversal: the ordering and its reverse are the only
compatible orderings.

\noindent \textbf{Richness.} Each triple 
$x,y,z\in X$ appears in exactly one
$\displaystyle{\mathcal{B}_\mu}$-comparison.

\textit{Richness} requires that exactly one of $\B{x}{y}{z}$,
$\B{y}{x}{z}$, or $\B{z}{y}{x}$ holds for each triple of distinct
alternatives $x,y,z$.\@ This requirement applies to the model as a
whole: the choice behavior revealing betweenness may come from
different choice types for different triples.\@ Thus, no single choice
type needs to reveal betweenness for every triple.

\noindent \textbf{Negative Transitivity.} For every four distinct
alternatives $x,y,z,w\in X$, if $\B{y}{x}{z}$, then either
$\B{y}{x}{w}$ or $\B{y}{z}{w}$, but not both.

In the spirit of \textit{negative transitivity} \citep{kreps1988notes},
this condition requires that whenever $y$ lies
between $x$ and $z$, every other alternative $w$ lie on exactly
one side of $y$: either the $x$-side or the $z$-side.\@

\begin{theorem}[Revealed primitive ordering]
\label{revealed}
Let $\mu$ be a choice model whose associated betweenness relation
$\mathcal{B}_\mu$ satisfies \textit{richness}. Then
$\mathcal{B}_\mu$ satisfies \textit{negative transitivity} if
and only if there is a primitive ordering $>$ such that every choice
type in $\mu$ satisfies IRM and SAM according to $>$. Moreover,
whenever such an ordering exists, it is unique up to reversal.
\end{theorem}

\begin{proof}
Please see Section~\ref{localpf} in the Appendix.
\end{proof}

Under \textit{richness}, Theorem~\ref{revealed} characterizes when a primitive
ordering compatible with the choice model exists and shows that it is
identified up to reversal.\@ To the best of our knowledge, this
approach to identifying primitives from a choice model is novel.\@
In particular, if $\mu$ is the minimal self-ordered extension of the
rational choice model with respect to $>$, then $>$ and its reverse
are the only primitive orderings compatible with $\mu$.\@ We formalize
this observation in the following corollary.

\begin{corollary}
\label{uniquness}
Let $\mu$ be the minimal self-ordered extension of the rational choice
model with respect to a primitive ordering $>$, and let $>'$ be a
distinct primitive ordering. Then, $\mu$ is also the minimal
self-ordered extension of the rational choice model with respect to
$>'$ if and only if $>'$ is the reverse of $>$.
\end{corollary}

\begin{proof}
Please see Section~\ref{localpf} in the Appendix.
\end{proof}

 %%%%%%%%%%%%%%%%%%%%%%%%%%%%%%%%%%%%%%%%%%%%%%%%
%%%%%%%%%%%%%%%%%%%%%%%%%%%%%%%%%%%%%%%%%%%%%%%%%%%%

\vspace*{-1cm}
\section{Universally self-ordered choice models}
\label{Section_universal}
We next ask whether a choice model can yield unique ordered
representations regardless of how the primitive orderings are
specified.\@ We define this requirement as follows.

\begin{definition}
A choice model $\mu$ is \textbf{universally self-ordered} if it is
self-ordered with respect to every partial order $\vartriangleright$
obtained from a collection of primitive orderings
$\{>_S\}_{S\in\X}$.
\end{definition}

Proposition~\ref{universal} characterizes universally self-ordered
models through a condition that strengthens the lattice requirement
and can be checked directly from the choice types in the model.
To state this condition, we first define what it means for a choice
type to be a mixture of two other choice types. Let $c,c'\in\C$.\@
A choice type $c^*\in\C$ is a \textbf{mixture} of $c$ and $c'$ if
$c^*(S)\in\{c(S),c'(S)\}$ for every $S\in\X$.\@
Here, a mixture is a deterministic choice type, not a probability
distribution over types.

\begin{definition}
A choice model $\mu$ is \textbf{mixture-closed} if every mixture of
any two choice types $c,c'\in\mu$ also belongs to $\mu$.
\end{definition}

Mixture-closedness means that admissible choice types can be
recombined menu by menu. If two choice types belong to the model,
then any choice type that follows the first on some menus and the
second on the remaining menus must also belong to the model.
Thus, while join and meet combine two choice types by selecting,
respectively, the higher- and lower-ranked of their two choices according to the
primitive ordering on each menu, a mixture permits any
menu-by-menu recombination of their choices.

\begin{proposition}
\label{universal}
A choice model is universally self-ordered if and only if it is
mixture-closed.
\end{proposition}

\begin{proof}
Please see Section~\ref{universalpf} in the Appendix.
\end{proof}

Proposition~\ref{universal} shows how to extend a choice model to
make it universally self-ordered and highlights how demanding
this requirement is.\@
Consider a choice model $\mu$ consisting of two choice types
that maximize preference relations $\succ_1$ and $\succ_2$,
respectively.\@ For fixed primitive orderings, we can make $\mu$
self-ordered by adding at most two choice types.\@ To make it
universally self-ordered, however, we must include every choice
type that selects either the $\succ_1$-maximal or the
$\succ_2$-maximal alternative on each menu, allowing the selection
to vary freely across menus.\@

More generally, if the choice domain contains every menu, the
only universally self-ordered extension of the rational choice
model is the model containing all choice types.\@ By contrast,
for a fixed primitive ordering, Theorem~\ref{minimalextension}
shows that the minimal self-ordered extension consists precisely
of the choice types satisfying IRM and SAM.\@
Finally, Example~\ref{RMR} illustrates how the mixture-closedness
criterion in Proposition~\ref{universal} can be used to verify
that a choice model is universally self-ordered.

\begin{example}\citep*{KRS}\normalfont\label{RMR}
Let $\{\succ_k\}_{k=1}^{K}$ be a $K$-tuple of strict preference
relations on $X$.\@ Define $\mu$ as the set of choice types $c$
such that, for each $S\in\X$, the alternative $c(S)$ is the
$\succ_k$-maximal alternative in $S$ for some $k$, which may vary
across menus.\@
To see that $\mu$ is universally self-ordered, consider any
$c,c'\in\mu$ and any mixture $c^*$ of these choice types.
For each $S\in\X$, we have $c^*(S)\in\{c(S),c'(S)\}$.
Since each of these two alternatives is maximal in $S$ according
to at least one preference relation in $\{\succ_k\}_{k=1}^{K}$,
so is $c^*(S)$. Hence, $c^*\in\mu$.\@
Thus, $\mu$ is \textit{mixture-closed} and, by
Proposition~\ref{universal}, universally self-ordered.
\end{example}

\vspace{-1cm}
\section{Final comments}

This paper studies when ordered heterogeneity can be used to resolve the
underidentification of probabilistic choice without abandoning the behavioral
restrictions of a prespecified model. The key requirement is model-internal:
the unique ordered representation of any probabilistic choice behavior
explained by the model must itself use only choice types admitted by that
model. We call a model satisfying this property self-ordered.

Our main result shows that this requirement has an exact structural
characterization.  The characterization  provides a direct recipe for modifying models
that fail this requirement. Applying this recipe to rational choice, we
identify its minimal self-ordered extension and show that the resulting model
permits disciplined departures from rational choice. A second contribution concerns the ordering itself. Rather than treating the
primitive ordering as necessarily exogenous, we show that it can be recovered
from the restrictions embodied in the choice model. 

These results suggest two natural directions for future work. One is to
characterize minimal self-ordered extensions of other economically relevant
behavioral models and to study the interpretation of the additional types
required for lattice closure. Another is to investigate how different
economically motivated primitive orderings affect the recovered distribution
of types and the  conclusions drawn from that distribution.

\newpage
%In this vein, we observe that   they are in a one-to-one correspondence %with   lattices.\@ It follows that  
 %self-ordered models  allow for a generality that is economically significant. %Consequently, self-ordered models offer a level of generality that holds %economic significance.\@ 

%[search for models violations of rational choice paradigm analyze/extend/explore/refine/structure % boundedly rational choice models tailored  for extensions of   ]

\bibliographystyle{agsm}
%\bibliographystyle{te}

%\newpage
%\bibliography{10_PRC}

\bibliography{choicematchingcombined}

@article{HuberPaynePuto1982,
  author  = {Huber, Joel and Payne, John W. and Puto, Christopher},
  title   = {Adding Asymmetrically Dominated Alternatives:
             Violations of Regularity and the Similarity Hypothesis},
  journal = {Journal of Consumer Research},
  year    = {1982},
  volume  = {9},
  number  = {1},
  pages   = {90--98},
  doi     = {10.1086/208899}
}

@article{HuberPuto1983,
  author  = {Huber, Joel and Puto, Christopher},
  title   = {Market Boundaries and Product Choice:
             Illustrating Attraction and Substitution Effects},
  journal = {Journal of Consumer Research},
  year    = {1983},
  volume  = {10},
  number  = {1},
  pages   = {31--44},
  doi     = {10.1086/208943}
}

@article{RieskampBusemeyerMellers2006,
  author  = {Rieskamp, J{\"o}rg and Busemeyer, Jerome R.
             and Mellers, Barbara A.},
  title   = {Extending the Bounds of Rationality:
             Evidence and Theories of Preferential Choice},
  journal = {Journal of Economic Literature},
  year    = {2006},
  volume  = {44},
  number  = {3},
  pages   = {631--661},
  doi     = {10.1257/jel.44.3.631}
}

@book{knuth1976marriages,
  title = {Mariages Stables},
  author = {Knuth, Donald E.},
  year = {1976},
  publisher = {Les Presses de l'Universit\'e de Montr\'eal},
  address = {Montr\'eal}
}

@book{rothsotomayor1990,
  title = {Two-Sided Matching: A Study in Game-Theoretic Modeling and Analysis},
  author = {Roth, Alvin E. and Sotomayor, Marilda A. O.},
  year = {1990},
  publisher = {Cambridge University Press},
  address = {Cambridge}
}

@article{dogan2015efficient,
  title = {Efficient and Stable Random Assignments in Marriage Problems},
  author = {Do\u{g}an, Battal and Y{\i}ld{\i}z, Kemal},
  journal = {Games and Economic Behavior},
  volume = {90},
  pages = {210--227},
  year = {2015}
}

@article{murota2016discrete,
  title={Discrete convex analysis: A tool for economics and game theory},
  author={Murota, Kazuo},
  journal={The Journal of Mechanism and Institution Design},
  volume={1},
  number={1},
  pages={151--273},
  year={2016},
  publisher={Society for the Promotion of Mechanism and Institution Design, University of~…}
}

@article{ausubel2006efficient,
  title={An efficient dynamic auction for heterogeneous commodities},
  author={Ausubel, Lawrence M},
  journal={American Economic Review},
  volume={96},
  number={3},
  pages={602--629},
  year={2006},
  publisher={American Economic Association}
}

@article{GulStacchetti1999,
  author  = {Gul, Faruk and Stacchetti, Ennio},
  title   = {Walrasian Equilibrium with Gross Substitutes},
  journal = {Journal of Economic Theory},
  year    = {1999},
  volume  = {87},
  number  = {1},
  pages   = {95--124},
  doi     = {10.1006/jeth.1999.2521}
}

@unpublished{Veinott1989,
  author = {Veinott, Arthur F.},
  title  = {Lattice Programming},
  note   = {Unpublished lecture notes, Johns Hopkins University},
  year   = {1989}
}

@book{Topkis1998,
  author    = {Topkis, Donald M.},
  title     = {Supermodularity and Complementarity},
  publisher = {Princeton University Press},
  address   = {Princeton, NJ},
  year      = {1998}
}

@article{MilgromShannon1994,
  author  = {Milgrom, Paul and Shannon, Chris},
  title   = {Monotone Comparative Statics},
  journal = {Econometrica},
  volume  = {62},
  number  = {1},
  pages   = {157--180},
  year    = {1994},
  month   = {January},
  publisher = {Econometric Society}
}

@article{juarez2022lattice,
  title={Lattice structure of the random stable set in many-to-many matching markets},
  author={Juarez, Noelia and Neme, Pablo and Oviedo, Jorge},
  journal={Games and Economic Behavior},
  volume={132},
  pages={255--273},
  year={2022},
  publisher={Elsevier}
}

@Article{topkis1978minimizing,
  author    = {Topkis, Donald M},
  title     = {Minimizing a submodular function on a lattice},
  journal   = {Operations Research},
  year      = {1978},
  volume    = {26},
  number    = {2},
  pages     = {305--321},
  publisher = {INFORMS},
}

@techreport{apesteguia2023testable,
  title={Testable implications of parametric assumptions in ordered random utility models},
  author={Apesteguia, JOSE and Ballester, MIGUEL A},
  year={2023},
  institution={Working paper}
}

@Article{KRS,
  author    = {Kalai, Gil and Rubinstein, Ariel and Spiegler, Ran},
  title     = {Rationalizing choice functions by multiple rationales},
  journal   = {Econometrica},
  year      = {2002},
  volume    = {70},
  number    = {6},
  pages     = {2481--2488},
  publisher = {Wiley Online Library},
}

@Article{scrum,
  author    = {Apesteguia, Jose and Ballester, Miguel A and Lu, Jay},
  title     = {Single-Crossing Random Utility Models},
  journal   = {Econometrica},
  year      = {2017},
  volume    = {85},
  number    = {2},
  pages     = {661--674},
  publisher = {Wiley Online Library},
}

@Article{prc,
  author    = {Filiz-Ozbay, Emel and Masatlioglu, Yusufcan},
  title     = {Progressive random choice},
  number    = {3},
  pages     = {716--750},
  volume    = {131},
  journal   = {Journal of Political Economy},
  publisher = {The University of Chicago Press Chicago, IL},
  year      = {2023},
}

@Article{similarity,
  author    = {Rubinstein, Ariel},
  title     = {Similarity and decision-making under risk: Is there a utility theory resolution to the {A}llais paradox?},
  number    = {1},
  pages     = {145--153},
  volume    = {46},
  journal   = {Journal of Economic Theory},
  publisher = {Elsevier},
  year      = {1988},
}

@Article{masatlioglu2020willpower,
  author    = {Masatlioglu, Yusufcan and Nakajima, Daisuke and Ozdenoren, Emre},
  title     = {Willpower and compromise effect},
  journal   = {Theoretical Economics},
  year      = {2020},
  volume    = {15},
  number    = {1},
  pages     = {279--317},
  publisher = {Wiley Online Library},
}

@Unpublished{petri2023ordered,
  author = {Petri, Henrik},
  title  = {Characterizing ordered models of behavioral heterogeneity},
  note   = {Working paper},
  year   = {2024},
}

@Article{fishburn1971betweenness,
  author    = {Fishburn, Peter C},
  title     = {Betweenness, orders and interval graphs},
  number    = {2},
  pages     = {159--178},
  volume    = {1},
  journal   = {Journal of Pure and Applied Algebra},
  publisher = {Elsevier},
  year      = {1971},
}

@Article{biro2023helly,
  author    = {Bir{\'o}, Csaba and Lehel, Jen{\H{o}} and T{\'o}th, G{\'e}za},
  title     = {Helly-Type Theorems for the Ordering of the Vertices of a Hypergraph},
  number    = {3},
  pages     = {665--682},
  volume    = {40},
  journal   = {Order},
  publisher = {Springer},
  year      = {2023},
}

@Article{huntington1924new,
  author  = {Huntington, Edward V},
  title   = {A new set of postulates for betweenness, with proof of complete independence},
  number  = {2},
  pages   = {257--282},
  volume  = {26},
  journal = {Transactions of the American Mathematical Society},
  year    = {1924},
}

@Article{huntington1917sets,
  author  = {Huntington, Edward V and Kline, J Robert},
  title   = {Sets of independent postulates for betweenness},
  number  = {3},
  pages   = {301--325},
  volume  = {18},
  journal = {Transactions of the American Mathematical Society},
  year    = {1917},
}

@Book{kreps1988notes,
  author    = {Kreps, David M},
  title     = {Notes on the theory of choice},
  publisher = {Westview Press},
  journal   = {Taylor \& Francis},
  year      = {1988},
}

@Article{costa2020single,
  author    = {Costa, Matheus and Ramos, Paulo Henrique and Riella, Gil},
  title     = {Single-crossing choice correspondences},
  number    = {1},
  pages     = {69--86},
  volume    = {54},
  journal   = {Social Choice and Welfare},
  publisher = {Springer},
  year      = {2020},
}

@Article{amemiya1981qualitative,
  author    = {Amemiya, Takeshi},
  title     = {Qualitative response models: A survey},
  number    = {4},
  pages     = {1483--1536},
  volume    = {19},
  journal   = {Journal of {E}conomic {L}iterature},
  publisher = {JSTOR},
  year      = {1981},
}

@Article{small1987discrete,
  author  = {Small, Kenneth A},
  title   = {A discrete choice model for ordered alternatives},
  journal = {Econometrica},
  volume  = {55},
  number  = {2},
  pages   = {409--424},
  year    = {1987},
}

@Book{Agresti1984,
  author    = {Agresti, Alan},
  title     = {Analysis of ordinal categorical data},
  publisher = {Wiley, New York.},
  year      = {1984},
}

@Article{barseghyan2021discrete,
  author    = {Barseghyan, Levon and Molinari, Francesca and Thirkettle, Matthew},
  title     = {Discrete choice under risk with limited consideration},
  number    = {6},
  pages     = {1972--2006},
  volume    = {111},
  journal   = {American Economic Review},
  publisher = {American Economic Association},
  year      = {2021},
}

@Article{tserenjigmid2021order,
  author    = {Tserenjigmid, Gerelt},
  title     = {The order-dependent {L}uce model},
  number    = {11},
  pages     = {6915--6933},
  volume    = {67},
  journal   = {Management Science},
  publisher = {INFORMS},
  year      = {2021},
}

@Article{turansick2022,
  author    = {Turansick, Christopher},
  title     = {Identification in the random utility model},
  pages     = {105489},
  volume    = {203},
  journal   = {Journal of Economic Theory},
  publisher = {Elsevier},
  year      = {2022},
}

@Article{petri2023binary,
  author    = {Petri, Henrik},
  title     = {Binary single-crossing random utility models},
  pages     = {311--320},
  volume    = {138},
  journal   = {Games and Economic Behavior},
  publisher = {Elsevier},
  year      = {2023},
}

@Article{masatlioglu2024growing,
  author    = {Masatlioglu, Yusufcan and Vu, Tri Phu},
  title     = {Growing attention},
  pages     = {105929},
  volume    = {222},
  journal   = {Journal of Economic Theory},
  publisher = {Elsevier},
  year      = {2024},
}

@Article{apesteguia2023random,
  author    = {Apesteguia, Jose and Ballester, Miguel A},
  title     = {Random utility models with ordered types and domains},
  pages     = {105674},
  volume    = {211},
  journal   = {Journal of Economic Theory},
  publisher = {Elsevier},
  year      = {2023},
}

@Article{cmy,
  author  = {Christopher P. Chambers and Yusufcan Masatlioglu and Kemal Yildiz},
  title   = {Ordered Probabilistic Choice},
  journal = {mimeo},
  year    = {2024},
}

@InProceedings{ganter1998stepwise,
  author       = {Ganter, Bernhard and Kuznetsov, Sergei O},
  booktitle    = {International Conference on Conceptual Structures},
  title        = {Stepwise construction of the {D}edekind-{M}acNeille completion},
  organization = {Springer},
  pages        = {295--302},
  year         = {1998},
}

@Article{nourine1999fast,
  author    = {Nourine, Lhouari and Raynaud, Olivier},
  title     = {A fast algorithm for building lattices},
  number    = {5-6},
  pages     = {199--204},
  volume    = {71},
  journal   = {Information processing letters},
  publisher = {Elsevier},
  year      = {1999},
}
\newpage 

\section{Appendix}
\label{appendix}
%\subsection{Lemmas for Theorem \ref{submodular}} 
%We present and prove the 

\subsection{Proof of Theorem \ref{minimalextension}}
\label{minimalproof}
Let $\mu^*$ denote the minimal self-ordered extension of the rational choice model with respect to $\vartriangleright$, and let $\nu$ denote the choice model consisting of the choice types that satisfy IRM and SAM. Since $\mu^*$ is self-ordered, Theorem \ref{progressiveifflattice} implies that $\langle \mu^*, \vartriangleright \rangle$ is a lattice and that there is no $\mu\subsetneq \mu^*$ that contains every rational choice type and for which $\langle \mu, \vartriangleright \rangle$ is a lattice.

We first show that $\mu^*\subset\nu$.\@ To see this, first note that every rational choice type $c$ belongs to $\nu$, since for each $S\in \X$ and $x\in S$, rationality of $c$ implies that $c(S)\neq c(S\setminus \{x\})$
only if $x=c(S)$.
%if $c(S) > x$  or  $x > c(S) $, then rationality of $c$ implies that  $c(S\setminus \{x\})=c(S)$, thus   $\succ_{S}\n=\succ_{S\setminus \{x\}}$.
Next,  we show that $\langle \nu, \vartriangleright \rangle$ is a lattice.\@
Let $c^1, c^2\in \nu$ and   $c=c^1\vee c^2$. Then,
to see that $c$ satisfies IRM and SAM, assume w.l.o.g.\@ that $c(S) =c^1(S)$.\@ Now, if $c^1(S) > x$ then, since $c^1$ satisfies IRM, we have
$c^1(S\setminus \{x\}) \geq c^1(S)$.\@ It follows that $c(S\setminus \{x\})\geq c(S)$.\@
If $x > c^1(S)$, then $x > c^2(S)$.  Since $c^1$ and $c^2$ satisfy SAM, we have   $c(S)\geq c (S\setminus \{x\})$. Thus, we conclude that $c^1\vee c^2\in \nu$. Symmetric arguments show  that $c^1\wedge c^2\in \nu$ as well. Therefore, $\nu$ is self-ordered and contains every rational choice type, so the minimality of $\mu^*$ implies that $\mu^*\subset\nu$.

%  $\succ^1_{S\setminus x}$ and $\succ^2_{S\setminus \{x\}}$ are  more  aligned % with $>$  than   $\succ^1_{S}$. 
%It follows that 

Next, we show that $\nu\subset\mu^*$.\@ To see this, let $c\in \nu$.\@ Since $\langle\mu^*, \vartriangleright \rangle$ is a lattice, by Lemma \ref{midlemma}, it suffices to show that for each $\displaystyle{S,S'\in \X}$, there exists $c^*\in \mu^*$ such that $c^*(S)=c(S)$ and $c^*(S')=c(S')$. Let $\displaystyle{S,S'\in \X}$ such that $c(S)=a$ and $c(S')=a'$.
If $a=a'$, then $c(S)$ and $c(S')$ are obtained by maximizing a preference relation that ranks $a$ first.\@  If $a\neq a'$, then  assume w.l.o.g.\@ that $a> a'$.\@ Now, there are   two cases.
 
\noindent Case 1: Suppose that  $\{a,a'\}\not \subset S\cap S'$.\@ Then, let $c_1$ be a choice type maximizing a preference relation that ranks $a$ first and $a'$ second, and $c_2$ be a choice type maximizing a preference relation that ranks $a'$ first and $a$ second. If $a\notin S'$, set $c^*=c_1 \vee c_2$; if $a'\notin S$, set $c^*=c_1 \wedge c_2$. In both cases, $c^*(S)=a$ and $c^*(S')=a'$, and $c^*\in \mu^*$ since $\langle\mu^*, \vartriangleright \rangle$ is a lattice containing every rational choice type.

\noindent Case 2: Suppose that $\{a,a'\}\subset S\cap S'$.\@ 
 We first show that  either (i) there exists $x\in S\setminus S'$ with $x> a$  or (ii) there exists $y\in S'\setminus S$ with $a' > y$. 
If not, then consider $S\cap S'$. Suppose that we remove the alternatives $x\in S\setminus S'$ from $S$ one by one.\@ Since $c\in \nu$, by applying IRM at each step, we get $c(S\cap S') \geq c(S)$.\@ Similarly, suppose that we remove the alternatives $y\in S'\setminus S$ from $S'$ one by one.\@ Then, by applying SAM, we get $c(S') \geq c(S\cap S')$.\@ Therefore,   we  have  $a'\geq a$, a contradiction.\@ Thus,  (i) or (ii) holds. Suppose that (i) holds.\@ Then,  let $c^*=c_1 \wedge c_2$, where  $c_1$ maximizes a preference relation that ranks $x$ first and $a'$ second, and $c_2$  maximizes a preference relation that ranks $a$ first.\@ Suppose that (ii) holds. Then, let  $c^*=c_1 \vee c_2$, where $c_1$  maximizes a preference relation that ranks $y$ first and $a$ second, and $c_2$  maximizes a preference relation that ranks $a'$ first. In both cases, $c^*(S)=a$ and $\displaystyle{c^*(S')=a'}$, and $c^*\in \mu^*$ since $\langle\mu^*, \vartriangleright \rangle$ is a lattice containing every rational choice type.

%%%%%%%%%%%%%%%%%%%%%%%%%%%%%%%%%%%%%%%%%%%%%%%%%%%%%%%%%%%%%%%%
% APPENDIX: PROOFS OF THEOREM 3 AND COROLLARY 1
%%%%%%%%%%%%%%%%%%%%%%%%%%%%%%%%%%%%%%%%%%%%%%%%%%%%%%%%%%%%%%%%

\subsection{Proof of Theorem \ref{revealed} and Corollary \ref{uniquness}}
\label{localpf}

It is critical to identify an ordering that \textbf{agrees with the
betweenness relation}: for each triple of distinct alternatives
$x,y,z\in X$, if $\B{y}{x}{z}$, then $x>y>z$ or $x<y<z$.\@
We call such an ordering an \emph{agreeing ordering}.
Betweenness is a ternary relation whose study stems from its use in
axiomatizations of geometry.\@ For example,
\cite{huntington1917sets} proposed eleven different sets of axioms
characterizing the usual betweenness relation on the real line, most
of which can be adapted to replace \textit{richness} and \textit{negative transitivity}
jointly.\@ Our \textit{richness} condition appears almost directly in these
axiomatizations, whereas \textit{negative transitivity} is most similar to the
axioms used in more succinct characterizations provided by
\cite{huntington1924new} and \cite{fishburn1971betweenness}.%
\footnote{See axiom 9 used by \cite{huntington1924new} and axiom
$A3$ used by \cite{fishburn1971betweenness}.}

To prove Theorem~\ref{revealed}, we use a recent result by
\cite*{biro2023helly}, who provide a unified view of existing results.\@
They show that, for a betweenness relation specifying exactly one
middle alternative for every triple, if there is an agreeing ordering
for every four-element subset, then there is an agreeing ordering for the
whole set.\@\footnote{This result is in line with the fact that none
of the axioms in the aforementioned characterizations uses more than
four elements.} \textit{Richness} ensures that every triple is so specified.
To apply their result, we show in Lemma~\ref{local} that \textit{richness}
and \textit{negative transitivity} suffice for the existence of orderings
that ``locally'' agree with our betweenness relation.\@

\begin{lemma}
\label{local}
Let $\mu$ be a choice model such that the associated betweenness
relation $\mathcal{B}_\mu$ satisfies \textit{richness} and
\textit{negative transitivity}.\@ Then, for each distinct
$x,y,z,w\in X$, there is an ordering $>_L$ such that for each triple
of distinct alternatives $a,b,c\in\{x,y,z,w\}$, if $\B{b}{a}{c}$,
then $a>_L b>_L c$ or $a<_L b<_L c$.
\end{lemma}

\begin{proof}
By \textit{richness}, assume w.l.o.g.\@ that $\B{y}{x}{z}$.\@
It follows from \textit{negative transitivity} that $\B{y}{x}{w}$ or
$\B{y}{z}{w}$, but not both.\@ Interchanging $x$ and $z$ if
necessary, suppose that $\B{y}{x}{w}$. Since $\B{y}{z}{w}$ does
not hold, \textit{richness} implies that either $\B{z}{y}{w}$ or
$\B{w}{y}{z}$.

Suppose first that $\B{z}{y}{w}$.\@ Applying \textit{negative transitivity}
with $z$ as the middle element implies that $\B{z}{y}{x}$ or
$\B{z}{w}{x}$, but not both.\@ \textit{Richness} rules out $\B{z}{y}{x}$
because $\B{y}{x}{z}$. Therefore, $\B{z}{x}{w}$.\@ Thus, the
ordering $x>_L y>_L z>_L w$ agrees with all four comparisons on
$\{x,y,z,w\}$.

The case in which $\B{w}{y}{z}$ is symmetric: \textit{negative transitivity}
and \textit{richness} imply $\B{w}{x}{z}$, and the ordering
$x>_L y>_L w>_L z$ agrees with all four comparisons.\@
\end{proof}

\begin{proof}[Proof of Theorem \ref{revealed}]
Suppose first that there is a primitive ordering $>$ such that every
choice type in $\mu$ satisfies IRM and SAM according to $>$.\@
Then $>$ agrees with $\mathcal{B}_\mu$. Since \textit{richness} holds,
for each triple of distinct alternatives $x,y,z\in X$, we have
$\B{y}{x}{z}$ if and only if $x>y>z$ or $x<y<z$.\@
\textit{Negative transitivity} follows immediately from this equivalence.

Conversely, suppose that $\mathcal{B}_\mu$ satisfies \textit{richness} and
\textit{negative transitivity}.\@ If $|X|\leq 3$, \textit{richness} directly yields
an agreeing ordering. If $|X|\geq 4$, it follows from our
Lemma~\ref{local} and Theorem 1 by \cite*{biro2023helly}, applied
to the complete system of triples specified by \textit{richness}, that there
is an ordering $>$ over $X$ such that, for each triple of distinct
alternatives $x,y,z\in X$, if $\B{y}{x}{z}$, then $x>y>z$ or
$x<y<z$.\@
To see that every choice type in $\mu$ satisfies IRM and SAM
according to $>$, let $c\in\mu$ and consider removing an unchosen
alternative $z$ from a menu $S$. If the choice changes from $y$ to
$x$, then $\B{y}{x}{z}$. Thus, if $y>z$, we must have $x>y$,
as required by IRM; if $z>y$, we must have $y>x$, as required by
SAM. Both axioms hold trivially when the choice does not change.

Finally, since \textit{richness} holds, every agreeing ordering has exactly
$\mathcal{B}_\mu$ as its betweenness relation.\@ Such a relation
determines a linear ordering up to reversal: its endpoints are the
alternatives that never appear as a middle alternative, and, once
one endpoint is designated as highest-ranked, comparisons involving
that endpoint determine the relative ranking of every other pair.
The case $|X|\leq 2$ is immediate.\@ Therefore, $>$ and its reverse
are the only primitive orderings compatible with $\mu$.
\end{proof}

\begin{proof}[Proof of Corollary \ref{uniquness}]
Suppose that $\mu$ is the minimal self-ordered extension of the
rational choice model with respect to a primitive ordering $>$.\@
We show that $\mathcal{B}_\mu$ satisfies \textit{richness}.\@ To see this,
let $x,y,z\in X$ be a triple such that $x>y>z$. Consider the choice
type $c=c_1\vee c_2$, where $c_1$ maximizes a preference relation
that ranks $z$ first and $x$ second, and $c_2$ maximizes a preference
relation that ranks $y$ first.\@ Since
$\langle\mu,\vartriangleright\rangle$ is a lattice containing every
rational choice type, $c\in\mu$. Moreover,
$c(\{x,y,z\})=y$ and $c(\{x,y\})=x$, so $\B{y}{x}{z}$.\@
Every triple therefore appears in at least one
$\mathcal{B}_\mu$-comparison. Since every choice type in $\mu$
satisfies IRM and SAM according to $>$, no triple can appear in
more than one such comparison. Thus, $\mathcal{B}_\mu$ satisfies
\textit{richness}, and Theorem~\ref{revealed} implies that $>$ and its
reverse are the only compatible primitive orderings.
Conversely, reversing $>$ interchanges IRM and SAM, so
Theorem~\ref{minimalextension} implies that the minimal
self-ordered extension is unchanged.\@
\end{proof}

%%%%%%%%%%%%%%%%%%%%%%%%%%%%%%

%\footnote{Same observation  follows from Theorem 4 by  \cite{fishburn1971betweenness}.}

%As for the proof of item ii, suppose the $\mathcal{B_\mu}$ satisfies $B1$ and $sB2$.\@ Then, it follows from Theorem 4 by \cite{fishburn1971betweenness} that there exists a linear order $>$ over $X$ such that for each triple $x,y,z\in X$, we have $\B{y}{x}{z}$ if and only if $x>y>z$ or $x<y<z$, which we denote by $x-y-z$. Thus, the only if part implies that every choice type in $\mu$ satisfies IRM and SAM according to $>$, while the if part implies that $>$ and its inverse are the only such orderings.
\subsection{Proof of Proposition \ref{universal}}
\label{universalpf}

\begin{proof}
Only if. Suppose, for a contradiction, that there exist $c_1,c_2\in \mu$ and a choice type $c\notin \mu$ such that $c(S)\in \{c_1(S), c_2(S)\}$ for every $S\in \X$.\@ Then, for each $S\in \X$, define the primitive ordering $>_S$ such that $c(S)$ is highest-ranked.\@ Thus, $c=c_1\vee c_2$, so $\langle \mu, \vartriangleright \rangle$ is not a lattice.\@ By Theorem \ref{progressiveifflattice}, $\mu$ is not self-ordered with respect to this collection of primitive orderings, contradicting the assumption that $\mu$ is universally self-ordered.

\noindent  If.  Since the join and meet of any two choice types are mixtures of those types,     $\langle \mu, \vartriangleright\rangle$   is a lattice for any partial order $\vartriangleright$   obtained from a set of primitive orderings.\@ Then, it follows from Theorem \ref{progressiveifflattice} that $\mu$ is universally self-ordered.
\end{proof}

\end{document}